\documentclass[11pt,a4paper]{article}
\usepackage{jheppub}
\usepackage{amsmath}
\usepackage{multicol,bbm}
\usepackage{enumerate}
\usepackage{bibentry}

\graphicspath{{./Figs/}}

\newcommand{\eq}{\begin{equation}}
\newcommand{\eqe}{\end{equation}}

\newcommand{\la}{\langle}
\newcommand{\ra}{\rangle}
\newcommand{\bea}{\begin{eqnarray}}
\newcommand{\eea}{\end{eqnarray}}
\newcommand{\eqa}{\begin{eqnarray}}
\newcommand{\eqae}{\end{eqnarray}}

\def\be{\begin{equation}}
\def\ee{\end{equation}}
\def\ba{\begin{eqnarray}}
\def\ea{\end{eqnarray}}
\def\bea{\begin{eqnarray}}
\def\eea{\end{eqnarray}}
\def\eq{\begin{equation}}
\def\eqe{\end{equation}}
\def\eqa{\begin{eqnarray}}
\def\eqae{\end{eqnarray}}
\def\la{\langle}
\def\ra{\rangle}

\def\nl{\nonumber\\}

\def\ZZ{\mathcal{Z}}

\def\NN{\mathcal{N}}

\def\l{\langle}
\def\r{\rangle}

\def\one{\mbox{1 \kern-.59em {\rm l}}}

\def\cA{{\cal A}}

\def\cM{{\cal M}}

\title{More on Soft Theorems: Trees, Loops and Strings}
\author[a]{Massimo Bianchi}
\author[b,c]{Song He}
\author[b,d]{ Yu-tin Huang}
\author[e]{Congkao Wen}

\affiliation[a]{Dipartimento di Fisica, Universit\`{a} di Roma ``Tor Vergata"\\
\& I.N.F.N. Sezione di Roma ``Tor Vergata"
Via della Ricerca Scientifica, 00133 Roma, Italy}

\affiliation[b]{School of Natural Sciences, Institute for Advanced
Study, Princeton, NJ 08540, USA}

\affiliation[c]{Perimeter Institute for Theoretical Physics, Waterloo, ON N2L 2Y5,
Canada}

\affiliation[d]{Department of Physics and Astronomy, National Taiwan University, Taipei 10617, Taiwan, ROC} 

\affiliation[e]{Centre for Research in String Theory, Department of Physics, Queen Mary University of London, Mile End Road, London E1 4NS, UK}

\emailAdd{massimo.bianchi@roma2.infn.it, songhe@ias.edu, yutin@ias.edu, c.wen@qmul.ac.uk} 

\abstract{We study soft theorems in a broader context, addressing their fate at loop level and their universality in effective field theories and string theory. We argue that for gauge theories in the planar limit, loop-level soft gluon theorems can be made manifest already at the {\it integrand} level. In particular, we show that the planar integrand for $\mathcal{N}=4$ SYM satisfies the tree-level soft theorem to all orders in perturbation theory and provide strong evidence to this effect for integrands in $\mathcal{N}<4$ SYM. We consider soft theorems for non-supersymmetric Yang-Mills theories and gravity, and show the validity of integrand soft theorem, while loop corrections to the integrated soft theorems are intimately tied to the presence of conformal anomalies. We then address the question of universality of the soft theorems for various theories. In effective field theories with $F^3$ and $R^3$ interactions, the soft theorems are not modified. However for gravity theories with $R^2 \phi$ interactions, the sub-sub-leading order soft graviton theorem, which is beyond what is implied by the extended BMS symmetry, requires modifications at tree level for non-supersymmetric theories, and at loop level for $\mathcal{N}\leq4$ supergravity due to anomalies. Finally, for superstring amplitudes at finite $\alpha'$, via explicit calculation for lower-point examples as well as world-sheet OPE analysis for arbitrary multiplicity, we show that the superstring amplitudes satisfy the same soft theorem as its field-theory counterpart. This is no longer true for bosonic closed strings due to the presence of  $R^2 \phi$ interactions. }
\preprint{ROM2F/2014/04, }

\begin{document}
\maketitle

\section{Introduction} \label{sec:Intro}
It is well known that scattering amplitudes in gauge and gravity theories display universal behavior as one of the external leg becomes soft. Historically, soft theorems at tree level were derived using Feynman diagrams, at leading order~\cite{Weinberg}, and at sub-leading  orders for soft photons~\cite{Low, Low2}, and for soft gravitons~\cite{SoftGravy1}.  More recently soft theorems have been revived for gravity~\cite{Cachazo} and for Yang-Mills theory~\cite{Casali}, using BCFW recursion relations~\cite{Britto:2004ap, Britto:2005fq} for tree amplitudes\footnote{The sub-leading  soft graviton theorem was also proposed in \cite{SoftGravy2}. Both gauge and gravity soft theorems have been proven to hold in arbitrary dimensions~\cite{Schwab:2014xua, Afkhami-Jeddi:2014fia}, based on scattering equations~\cite{CHY}.}. One of the motivations for studying soft graviton theorems is to understand their relations with the conjectured new infinite dimensional symmetry of gravitational scattering amplitudes~\cite{Strominger1, Strominger2, Strominger3, Strominger4, Barnich1, Barnich2, Barnich3}, extending the Bondi, van der Burg, Metzner, and Sachs (BMS) symmetry~\cite{BMS} at null infinity. Given all these different ways of motivating and deriving soft theorems, it is natural to ask if these theorems are respected in more general gauge and gravity theories, including string theory. 

Furthermore, the soft behavior of loop-level amplitudes has been studied at leading order~\cite{BernSoft, KosowerSoft, BernAllPlus} and more recently at sub-leading orders~\cite{ZviScott, HHW}, for both gauge theories and gravity.  It is well known that the leading soft graviton theorem is protected from loop corrections~\cite{BernAllPlus}, but sub-leading soft graviton theorems and soft gluon theorems both require corrections at loop level. On the other hand, it has been argued in~\cite{FreddyEllis} that the distributional nature of the soft limit implies an alternative way of studying soft behaviors at loop level: one should first expand around the soft limit and then perform the loop integrals for the amplitude, which involves an expansion in the regulator. With this prescription, it has been shown in~\cite{FreddyEllis} that the sub-leading soft theorem is not renormalized in the example of one-loop five-point amplitude in $\mathcal{N}=8$ supergravity. Note that for the purpose of obtaining the correct infrared behavior for scattering amplitudes, it is necessary to abide by the usual procedure of regulating before taking the soft-limit~\cite{ZviScott}. The prescription prescribed by~\cite{FreddyEllis} instead serves as constraint one can impose on $D$-dimensional integrands. 

In this paper we will continue the investigation of soft theorems along these two directions: their fate at loop level, and their universality in effective field theories and string theory. First, we will examine loop-level soft theorem using the prescription of ~\cite{FreddyEllis}. In section \ref{integrand}, we will argue that for gauge theories in the planar limit, loop-level soft gluon theorems can be made manifest already at the {\it integrand} level. In particular, we will show that the planar integrands for $\mathcal{N}=4$ super Yang-Mills theory (SYM), determined by loop-level BCFW recursion relations~\cite{ArkaniHamed:2010kv}, satisfy the soft theorem to all loop orders, exactly as the tree amplitudes. For $\mathcal{N}<4$ SYM, we show explicitly that the same is true for one-loop MHV amplitudes in the CSW representation. In practice, our analysis is simplified significantly by using momentum-twistor variables \cite{Hodges:2009hk} and choosing to solve momentum conservation in a canonical way for the planar case.

For non-supersymmetric Yang-Mills theory or theories of gravity, no such representation of the integrands is known, thus one has to verify the soft theorems in the same way as in~\cite{FreddyEllis}, {\it i.e.} performing the integrals after the soft expansion of the integrand.  In section \ref{Allplus}, we will carefully examine the integrals from the soft expansion of all-plus one-loop integrands in both Yang-Mills and gravity, and show that both soft theorems are respected, {\it i.e.} the all-plus integrand has the interesting property that taking the soft-parameter and IR-regulator to zero in different orders commute. This is no longer the case for the single-minus amplitude as observed in~\cite{HHW}. For the latter, we demonstrate that the violation of tree-level soft theorem can be tied to the presence of conformal anomalies at loop level.

In addition to soft theorems at loop level, we also consider the question of how universal are they even at tree level. Naively one would expect that, sub-leading soft theorems may fail  in any effective field theory of gauge or gravity if the three-point interaction is modified. In section \ref{higher-dim}, we will study effective field theories with $F^3$ and $R^3$ interactions, and show that soft theorems are not altered in theses cases. A byproduct of our study is a BCFW recursion relation for $F^3$ amplitudes, written in momentum-twistor space in a form very similar to that of Yang-Mills amplitudes. However, for $R^2 \phi$ interactions, the sub-sub-leading soft graviton theorem needs modifications at tree level. Note that while such interactions can be suppressed at tree level via supersymmetry, they are generated in ${\cal N}\le 4$ supergravity due to the presence of $U(1)$ anomalies~\cite{RaduAnomaly}. This modification does not contradict with that implied by BMS symmetry, since the latter only predicts universality for the sub-leading soft behavior.

A more interesting aspect of universality is the soft theorems for tree-level string amplitudes. Although $\alpha'$-expansions of string amplitudes are coded in effective field theories, there is \textit{a priori} no Feynman-diagram-like argument for soft theorems at finite $\alpha'$. In section \ref{string1}, we will show, by explicit computations using four-dimensional kinematics for the cases of four and five points (six-point computation will be present in Appendix \ref{Appendix:sixpt}), that open superstring amplitudes on the disk satisfy the same soft gluon theorem as the corresponding gauge theory amplitudes. Using KLT relations~\cite{KLTref}, we will also verify the soft graviton theorem for four- and five-point closed superstring amplitudes. The above result can be understood via  BCFW recursion relations for string amplitudes. Combining BCFW recursion relations with the crucial observation that only massless states can contribute to the soft limit, we will argue generally that amplitudes for both bosonic and super open string theory satisfy the soft theorems. Whereas supersymmetric closed string theory satisfies the soft theorems, and sub-sub-leading term in soft theorem for bosonic closed-string amplitudes needs corrections.

Finally, we confirm the above analysis for general multiplicity from a world-sheet perspective. We will show that the soft behaviour is captured by the perator product expansion (OPE) of the soft vertex operator with adjacent vertex operators in the open string case and with any hard vertex operator in the closed string case. BRST symmetry will play a crucial role in the identification of the relevant terms in the OPE and in the choice of the picture for the colliding vertex operators. We will argue that soft theorems hold both in $D=10$ and in lower dimension where gauge boson and graviton vertex operator simply involve the identity operator of the CFT$_2$ governing the dynamics of the internal space. 
\\

{\bf Added Note}: In the completion of this manuscript, the work by Schwab~\cite{Schwab:2014fia} appeared on the arXiv which has some overlap with the results in section \ref{open45pt} and Appendix \ref{Appendix:sixpt}.

\subsection{Review}\label{sec:Review}

We begin with a brief review of soft theorems for tree-level amplitudes in gauge and gravity theories. The $n$-point amplitude involving the emission of a soft photon can be expanded in terms of the soft momentum $s$. The leading and sub-leading terms in this expansion are given by universal operators acting on the $(n{-}1)$-point amplitude, a fact that is well understood ever since the work of Low~\cite{Low} who recognized this as a simple consequence of gauge invariance. To see this, separate the Feynman diagrams into two classes:
$$\includegraphics[scale=0.4]{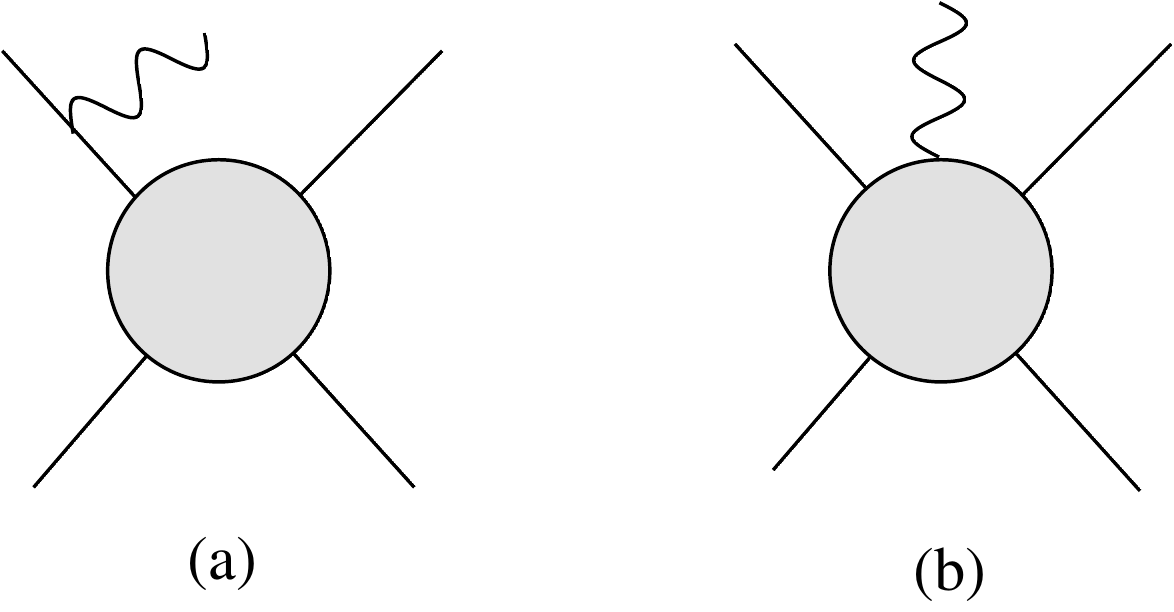}\,.$$

Diagram (a) has the soft photon connected to an external line which contributes to the leading divergence in the soft limit, proportional to  $\sum_{i}e_i(\epsilon{\cdot}k_i)/(s{\cdot}k_i)$ multiplied by the remaining hard amplitude with one leg slightly off-shell. Sub-leading terms are distributed between diagram (a) and (b) where the soft photon is connected to an internal line of the Feynman diagram. Using the fact that the sub-leading contribution from diagram (a) violates the Ward identity, which is generated by expanding the $(n{-}1)$-point amplitude near $s=0$, gauge invariance requires the sub-leading contribution from diagram (b) to be given by differential operators acting on the $(n{-}1)$-point amplitude. 

This observation allowed Low to express the sub-leading soft limit as a universal soft operator acting on the $(n{-}1)$-point amplitude. For further extension of Low's result see~\cite{Low2}. Generalizing Low's argument to gravity, Gross and Jackiw~\cite{SoftGravy1} obtained soft theorems for gravity accurate up to terms of order $\mathcal{O}(s^2)$, to be compared with $\mathcal{O}(s)$ for gauge theory. Thus the tree-level soft theorems for gravity is universal up to sub-sub-leading  in $s$. For a more recent analysis see~\cite{SoftGravy2}.  

An alternative way to derive the soft theorems is by using BCFW recursion relations for Yang-Mills and gravity, as was done in~\cite{Cachazo, Casali}. Consider the BCFW representation for tree-level gravity amplitude and choose the soft leg to be one of the shifted lines. If the soft graviton is plus helicity, shift the spinors holomorphically,
\bea \label{BCFWshift}
{\lambda}_{\hat{s}} = 
{\lambda}_{s} +  z \lambda_{n}\,,\quad \tilde{\lambda}_{\hat{n}} = 
\tilde{\lambda}_{{n}} -  z \tilde{\lambda}_{s}  \, ,
\eea
the BCFW representation is given by:
\eq
\label{BCFWrep}
M_{n+1}(1,2,\ldots,n, s^+)=\sum_{1\leq i<n}M_3(\hat{s}^+,i,-\hat{K}_{i s})\frac{1}{K^2_{i s}} M_{n}(\hat{K}_{i s},\ldots,\hat{n})+R\,,
\eqe
where $\hat{K}_{i s} = k_i + k_{\hat{s}}$, and $R$ represents terms arising from factorization poles $1/(k_s+K)^2$ with $K$ a non-null momentum. The holomorphic soft limit is achieved by scaling $\lambda_s\rightarrow \delta\lambda_s$. It was shown explicitly in~\cite{Cachazo} that the function $R$ is finite under the holomorphic soft limit, thus 
\eq\label{BCFWDiv}
M_{n+1}(1,2,\ldots,n,\{\delta\lambda_s,\tilde\lambda_s\}^+)\bigg|_{\rm div}=\sum_{1\leq i<n}M_3(\hat{s}^+,i,-\hat{K}_{i s})\frac{1}{K^2_{i s}} M_{n}(\hat{K}_{i s},\cdots,\hat{n})\bigg|_{\rm div}\,,
\eqe
where each term on the RHS can be written as:
\bea \nonumber
M_3(\hat{s}^+,i, -\hat{K}_{i s} )\frac{1}{K^2_{is}}M_{n}(\hat{K}_{i s},\ldots,\hat{n})= \mathcal{S}_{s i}
M_{n}( 
\{  \lambda_i, 
\tilde{\lambda}_i + \delta{ \langle s  n \rangle \over \langle i  n\rangle } \tilde{\lambda}_s \}   , 
\, \ldots \, , 
\{  \lambda_n, 
\tilde{\lambda}_n + \delta{ \langle s  i \rangle \over \langle n  i\rangle } \tilde{\lambda}_s \} ) \, , \\
\eea
where ``$\ldots$" indicate un-shifted $\{ \lambda\, , \tilde{\lambda} \}$, and $\mathcal{S}_{s i}$ is the ``inverse-soft-function" that is independent of the helicity of the $i$-th leg:
\eq
\mathcal{S}_{s i}={1 \over \delta^3} \frac{\langle ni\rangle^2 [i s] }{\langle ns\rangle^2 \langle i s\rangle}\,.
\eqe
Expanding $M_{n}(\hat{K}_{i s},\cdots,\hat{n})$ in $\delta$, it is straight forward to obtain the divergent part of the holomorphic soft-limit
\eqa\label{CSsoft}
 M_{n+1}(1,\ldots,n,\{ \delta\lambda_s, \tilde{\lambda}_s\}^+)\bigg|_{\rm div}&=&\left(\frac{1}{\delta^3}S_{\rm G}^{(0)}+\frac{1}{\delta^2}S_{\rm G}^{(1)}+\frac{1}{\delta}S_{\rm G}^{(2)}\right)M_{n}
\eqae
where the operator $S_{\rm G}^{(k)}$ is defined as:
\bea
S_{\rm G}^{(k)} = 
\sum^{n-1}_{i=1} {1 \over k! }\mathcal{S}_{s i}
\left({ \langle s  n \rangle \over \langle i  n\rangle } \tilde{\lambda}_s 
\cdot { \partial \over  \partial \tilde{\lambda}_{i} } +{ \langle s  i \rangle \over \langle n  i\rangle } \tilde{\lambda}_s 
\cdot { \partial \over  \partial \tilde{\lambda}_{n} } \right)^k  \, .
\eea
Note that here, $M_{n}$ is still subject to the $(n{+}1)$-pt amplitude momentum conservation, which is solved by expressing two $\tilde\lambda$'s in terms of the remaining $(n{-}1)$ ones. 

Now we turn to the soft gluon theorem. Throughout the paper, we will consider color-ordered, partial amplitudes for gluons (in any gauge theories and open-string theories):
\be~\label{color-decomposition}
{\bf A}_n(\{1^{a_1},2^{a_2},\ldots, n^{a_n}\}) = \sum_{\sigma\in  S_n/Z_n} {\rm Tr}(T^{a_{1_\sigma}} T^{a_{2_\sigma}} \cdots T^{a_{n_\sigma}})  A_n({1_\sigma}, {2_\sigma} \ldots {n_\sigma})
\ee
where ${\bf A}$ denotes the full, color-dressed amplitude and $A$ the corresponding color-ordered amplitude. This is the color-decomposition at tree level, but as we will restrict to gauge theories in the planar limit whereby $N_c\to \infty$, eq.~(\ref{color-decomposition}) applies to loop amplitudes as well. 

The soft gluon theorem can be derived in a parallel fashion with gravity by using the BCFW representation of tree-level color-ordered amplitudes: the divergent term in the holomorphic soft limit is again isolated into the two particle channel (only one term $i=1$ contributes because of the color ordering), and we find
\bea \label{BCFWYM}
&& A_{n+1}(  \{  \lambda_1, \tilde{\lambda}_1 \}\, , \ldots \,,  
\{  \lambda_n, \tilde{\lambda}_n \}\, , \{ \delta  \lambda_s,  \tilde{\lambda}_s \}^+ )\bigg|_{\rm div}
\cr
&=& \sum_{k=0,1} \frac{1}{\delta^{2-k}}S_{\rm YM}^{(k)} (n\,s\,1)
A_{n}(  \{  \lambda_1, \tilde{\lambda}_1 \}\, , \ldots \,,  
\{  \lambda_n, \tilde{\lambda}_n \}) 
\eea
with
\bea \label{SkYM}
S^{(k)}_{\rm YM} (n \, s\, 1)= 
{1 \over k!} { \langle n 1 \rangle  \over \langle n s\rangle \langle s 1\rangle  } 
\left( { \langle s  n \rangle \over \langle 1 n\rangle } \tilde{\lambda}_s \cdot { \partial \over \partial
 \tilde{\lambda}_1 } +
 { \langle s  1 \rangle \over \langle n  1\rangle } \tilde{\lambda}_s \cdot { \partial \over \partial
 \tilde{\lambda}_n }   \right)^k \, .
\eea
So for tree-level amplitudes in Yang-Mills theories, only $S^{(0)}_{\rm YM}$ and $S^{(1)}_{\rm YM}$ are universal. Note that if we choose to solve momentum conservation by expressing $\tilde{\lambda}_1, \tilde{\lambda}_n$ in terms of linear combinations of the remaining anti-holomorphic spinors, the sub-leading  soft terms actually vanishes! This prescription is more natural for planar amplitudes, especially when expressed using momentum twistors, as we will see shortly.

The derivation of the soft theorem from the recursion relation mirrors the work by Low, in that the contribution stems from two-particle channels that involve the soft leg. While in Low's work the sub-leading contribution also stems from diagrams where the soft leg is attached to an internal line, they are controlled by the leading contribution via Ward identities. Since the representation based on recursion relations uses gauge invariant building blocks, it is not a surprise that only the aforementioned two-particle channels contribute. 


\section{The soft gluon theorem for loop integrands in planar SYM}\label{integrand}

In this section, we consider supersymmetric Yang-Mills theories in the planar limit. The advantage of working with these theories is that one can determine the four-dimensional integrand at any loop order, as a rational function of external and loop momenta. We will argue that, the Yang-Mills soft theorem works directly at the level of the integrand, in essentially the same way as the BCFW derivation at tree level, which we just reviewed. As explained in~\cite{FreddyEllis}, if we interpret the soft theorem as a distributional relation, one needs to evaluate the loop integral (with suitable IR and UV regulators) after the soft expansion.  Given that we will show the tree-level soft theorem holds for the four-dimensional integrand, with the prescription of~\cite{FreddyEllis}, even after regularization and the expansion in terms of the regulator, we expect that the soft behavior of loop amplitudes is not renormalized. 

For color-ordered amplitudes in the planar limit, we find it convenient to choose the momenta adjacent to the soft particle for solving momentum conservation, in which case the soft theorem states that the sub-leading term should vanish. We will show that this is indeed the case for loop integrands of amplitudes in planar SYM theories with $\NN$ supercharges. For convenience, let us strip off an overall MHV pre-factor   
\be
A_0 \equiv \frac{\delta^{4|2 \NN} (\sum_{a=1}^n \lambda^{\alpha}_a (\tilde\lambda^{\dot{\alpha}}_a|\eta^A_a) )}{\l 1 2\r \ldots\l n{-}1 n\r\l n 1\r}, 
\ee
with $\alpha=1,2$, $\dot{\alpha}=\dot 1, \dot 2$ Lorentz indices, and $A=1,\ldots,\NN$ the SU$(\NN)$ R-symmetry index. Note that by definition, MHV tree amplitudes is given by $\l a, b\r^{4{-}\NN}$ where $a,b$ are the two negative-helicity particles (for $\NN=4$ it is simply unity).  

For the $n$-point, N${}^{k}$MHV amplitude at $L$ loops, $A_{n,k}^{(L)}$, let us denote the integrand (after stripping-off  $A_0$) by $R^{(L)}_{n,k}$: 
\be 
A_{n,k}^{(L)}=A_0 \times \int d^D \ell_1\cdots d^D \ell_L R^{(L)}_{n,k} (1,\cdots, n; \ell_1,\cdots,\ell_L), 
\ee 
where $\ell_1,\cdots \ell_L$ denotes the loop variables, and $D=4{-}2\epsilon$ with $\epsilon$ being the dimensional regulator. 

The soft limit of the planar integrand, including the tree amplitude for $L=0$, can be most conveniently written in terms of momentum twistor variables~\cite{Hodges:2009hk}. These are variables that trivialize momentum conservation and in terms of which spinor-helicity variables can be expressed as linear combinations. We write these (super) momentum twistors (with $4|\NN$ components) as $\ZZ_a=(Z^I_a|\eta^A_a)=(\lambda^{\alpha}_a, \mu^{\dot{\alpha}}_a|\chi^A_a)$ for $a=1,\ldots, n$, where for the bosonic part $Z^I_a$, the first two components are the holomorphic spinors $\lambda^{\alpha}$ and the remaining two components can be used to express the anti-holomorphic spinors $\tilde\lambda^{\dot{\alpha}}$ as follows:
\be
\tilde\lambda^{\dot{\alpha}}_a=\frac {\mu^{\dot{\alpha}}_{a{-}1}} {\l a \, a{+}1\r} +\frac{\l a{-}1 \, a{+}1\r \mu^{\dot{\alpha}}_a}{\l a{-}1 \, a\r \l a \, a{+}1\r} +\frac{\mu^{\dot{\alpha}}_{a{+}1}}{\l a{-}1 \, a\r}, ~\label{tlambda}
\ee   
for $a=1,\ldots, n$ with $a\pm 1$ modulo $n$. The Grassmann variables $\eta^A$ can be written as the same linear combination of the Grassmann part of the twistors $\chi^A$ as $\tilde\lambda^{\dot{\alpha}}$ of $\mu^{\dot{\alpha}}$.  In addition, loop variables are given by $L$ bi-twistors $\ell_1=(A_1, B_1),\ldots, \ell_L=(A_L, B_L)$. In terms of these variables,  $R^{(L)}_{n,k}$ is a degree-$(4k{-}8)$ polynomial of $\chi^A$'s and a rational function of the totally anti-symmetric contractions $\l a b c d\r\equiv \epsilon_{I J K L} Z^I_a Z^J_b Z^K_c Z^L_d$ of external and loop (bosonic) twistors. Note that the two-bracket of holomorphic spinors are given by $\l a b\r\equiv \l a b I\r$ where $I$ is the infinity (bi)twistor projecting any twistor to its first two components. 

We now consider the soft limit in momentum twistor space. Taking leg $n$ to be soft, we approach the soft limit by deforming 
\eq\label{TwistorSoft}
Z_n  \rightarrow \alpha Z_{n{-}1}+\beta Z_1+\delta Z_s
\eqe
where $\delta$ is the soft parameter. To see why this corresponds to the soft limit, from eq.~(\ref{tlambda}), observe that 
the deformation in eq.(\ref{TwistorSoft}) implies
\eq
\tilde{\lambda}_n=\delta\;\frac{\la n{-}1 1\ra \mu_s+\la 1 s\ra \mu_{n{-}1}+\la s n{-}1\ra \mu_{1}}{\langle 1 n{-}1 \ra^2\alpha\beta}
\eqe
Thus this limit corresponds to the anti-holomorphic soft limit. Furthermore, since $\tilde{\lambda}_a$ is determined by the twistors $(Z_{a{-}1},Z_{a},Z_{a{+}1})$, the deformation in eq.(\ref{TwistorSoft}) corresponds to deforming $\tilde{\lambda}_{n{-}1}$ and $\tilde{\lambda}_1$ as well, {\it i.e.} the momentum conservation is preserved by having all $a\neq (n{-}1, 1)$ $\tilde{\lambda}_a$'s fixed and solving $\tilde{\lambda}_{n{-}1}$ and $\tilde{\lambda}_1$ in terms of them. This is precisely the prescription that leads to vanishing sub-leading  soft corrections, as discussed in sec.\ref{sec:Review}, which can now be written in momentum-twistor space:
\be
n~~{\rm points}: \{Z_1, \ldots, Z_{n{-}1}, Z_n=\alpha Z_{n{-}1}{+}\beta Z_1{+}\delta Z_s \}; \quad (n{-}1)~~{\rm points}: \{Z_1, \ldots, Z_{n{-}1}\}. 
\ee

For the rest of the section, we would like to show that the sub-leading  soft expansion of momentum-twistor space integrand begins at $\mathcal{O}(\delta^0)$ for a negative-helicity soft leg, and at $\mathcal{O}(\delta^2)$ for a positive-helicity soft leg.\footnote{It is $\mathcal{O}(\delta^2)$ for the positive-helicity leg because we need to rescale the holomorphic soft behavior by $\delta^2$ to see the anti-holomorphic soft behavior.} It suffices to focus on the case of  a positive-helicity particle, {\it i.e.} the $k$-preserving soft limit, in which case we will take eq.~(\ref{TwistorSoft}) supersymmetrically.  Note that the MHV pre-factor absorbs the leading soft factor $S^{(0)}_{\rm YM}$, thus making the stripped amplitude behave trivially at leading order.  We claim that the following soft theorem holds for the planar integrand of SYM to any loop order:  
\be\label{integrandsoft}
R^{(L)}_{n,k}(\ZZ_1,\ldots, \ZZ_n)= R^{(L)}_{n{-}1,k}(\ZZ_1,\ldots,\ZZ_{n{-}1})+ 0\times \delta+ \mathcal{O}(\delta^2).
\ee
\subsection{All-loop integrand of $\NN=4$ SYM }
We first consider the $\mathcal{N}=4$ integrand, which satisfies a BCFW-like recursion relation most compactly written in momentum-twistor space~\cite{ArkaniHamed:2010kv},
\ba
\label{AHBCFWrecrel}
R^{(L)}_{n,k}&=&R^{(L)}_{n{-}1, k}+\sum_{L', k', i} R^{(L')}_{i, k'} (1,\cdots,i{-}1, I_i) [1, i{-}1, i, n{-}1, n] R^{(L{-}L')}_{n{+}2{-}i, k{-}1{-}k'} (I_i,i,\cdots,\hat{n}_i)\nl
+&&\int_{{\rm GL}(2)} [1, A, B, n{-}1, n] R^{(L{-}1)}_{n{+}2, k{+}1} (1,\cdots, \hat{n}, A, \hat{B}),
\ea
where we suppress the sum over distributions of loop variables $\ell_1, \ldots, \ell_L$ on both factorization and forward-limit terms, and for the latter one needs to perform fermionic and GL$(2)$ integrals. In addition, $\hat{n}_i=(n{-}1 n)\cap (1 i{-}1 i)$, $I_i=(i{-}1 i)\cap (1 n{-}1 n)$, $\hat{n}=(n{-}1 n)\cap (1 A B)$, $\hat{B}=(A B)\cap (1 n{-}1 n)$ with the intersection defined as $( a b)\cap (i j k)\equiv \ZZ_a \l b i j k\r-\ZZ_b \l a i j k\r$, and the R-invariant of five (super) twistors is defined as
\be
[a, b, c, d, e]\equiv \frac{\delta^{0|4}(\chi_a \l b c d e\r+{\rm cyc} )}{\l a b c d\r\l b c d e\r\l c d e a\r\l d e a b\r\l e a b c\r}. 
\ee

It is not a coincidence that we choose to shift the momentum-twistor $\ZZ_n$ of the soft particle \`a la BCFW. For this shift, the first term in the recursion corresponds to the special BCFW factorization term: the $(n{-}1)$-point, $k$-preserving amplitude, multiplied by three-point anti-MHV amplitude, and we will show that it is the only term that contributes to the first two orders of the soft expansion, which is a fact we are familiar with at tree level. This turns out to be a direct generalization of the BCFW argument for soft theorem at tree level. 

Let us first see how it works in this language for tree amplitudes, $L=0$, where only the first line contributes. In the soft limit, $I_i=\delta (i{-}1 i)\cap (1 n{-}1 s)\equiv \delta I'_i$, $\ZZ_{\hat{n}_i}=\ZZ_1{+} \mathcal{O}(\delta)$, thus the two sub-amplitudes are both non-singular as we take $\delta\to 0$. The R-invairant, $[1, i{-}1, i, n{-}1, n]$, however, becomes of order $\delta^2$:
\be\label{facR}
\frac{\delta^2}{\alpha\,\beta}\times \frac{\delta^{0|4} (\chi_{[i{-}1} \l i]\, n{-}1 \, s \, 1\r +\chi_s \l 1\, i{-}1 \, i \, n\r) }{\l 1\, i{-}1\, i\, n{-}1\r^3 \l  n{-}1\, s\, 1\, i{-}1 \r\l n{-}1\, s\, 1\, i\r}+\mathcal{O}(\delta^3)
\ee 
where in the numerator we have used the fact that terms involving $\chi_{n{-}1}$ and $\chi_1$ cancel with each other, and $[i{-}1, i]$ means antisymmetrization w.r.t. the two labels.  Thus we recovered the soft gluon theorem at tree level, 
\be
R^{(0)}_{n,k}=R^{(0)}_{n{-}1,k}+ \mathcal{O}(\delta^2),
\ee

Now it becomes clear that the first two orders in the soft expansion of the loop integrand are identical to those of tree amplitudes. The factorization part works exactly as before, except that now we need to use the fact that sub-amplitudes are non-singular at the loop integrand level.  For the forward-limit term, the R-invariant, $[1,A,B,n{-}1,n]$, behaves exactly as that in the factorization term,
\be\label{flR}
[1,A,B,n{-}1,n]=\frac{\delta^2}{\alpha\beta}\times \frac{\delta^{0|4} (\chi_{[A} \l B]\, n{-}1\, s\, 1\r +\chi_s \l 1\, A\, B\, n\r) }{\l A\, B\, 1\,n{-}1\r^3 \l 1\, A\, n{-}1\, s\r\l 1\, B\, n{-}1\, s\r}+\mathcal{O}(\delta^3) . 
\ee
In addition, the lower-loop integrand is again non-singular, with $\ZZ_{\hat{n}}=\ZZ_1{+}\mathcal{O}(\delta)$ and $\hat{B}=\delta (A B)\cap (1 n{-}1 s)\equiv \delta \hat{B}'$. After performing the fermionic and GL$(2)$ integrals we find that the entire forward-limit term goes like $\mathcal{O}(\delta^2)$ in the limit,  thus we conclude that  the soft-theorem holds for all-loop integrand in $\mathcal{N}=4$ SYM,
\be 
R^{(L)}_{n,k}=R^{(L)}_{n{-}1, k}+ \mathcal{O}(\delta^2). 
\ee
Note that although the sub-sub-leading ($\mathcal{O}(\delta^2)$) order is no longer universal, it takes a relatively simple form: it is given by factorization and forward-limit terms with eq.~(\ref{facR}), (\ref{flR}), where the dependence on the parameters is always through the prefactor $\delta^2/(\alpha \beta)$.

Before ending the discussion for $\NN=4$ SYM, let us look at the soft behavior of forward limit terms even more explicitly for the one-loop integrand. One can easily see that indeed each forward-limit term at one-loop goes like $\mathcal{O}(\delta^2)$ when we take the BCFW-shifted particle, $n$, to be soft. For example, forward-limit terms for one-loop MHV integrand, $K_{i,n}$ with $2<i<n$, are given by~\cite{ArkaniHamed:2010kv}:
\ba\label{kermit}
K_{i,n}&=-&\frac{\l A B (1 i{-}1 i )\cap (1 n{-}1 n)\r^2}{\l A B 1 i{-}1 \r\l A B 1 i \r\l A B i{-}1 i\r\l A B 1 n{-}1 \r\l A B 1 n \r\l A B n{-}1 n\r}\nl
&=&\frac{\delta^2}{\alpha\,\beta}\times \frac{ \l A B (1 i{-}1 i )\cap (1 n{-}1 s)\r^2}{\l A B 1 i{-}1 \r\l A B 1 i \r\l A B i{-}1 i\r\l A B 1 n{-}1 \r^3}+\mathcal{O}(\delta^3).\nl
\ea

\subsection{Integrands for $\NN<4$ SYM}
Now we turn to the soft theorem for $\NN<4$ SYM theories. It is illuminating to first write down BCFW recursion relations for tree amplitudes in any $\NN<4$ gauge theories, in terms of momentum-twistor variables~\cite{unpublished}, from which again the soft gluon theorem follows immediately. 


When taking the BCFW shift of $\ZZ_n$, without loss of generality we assume the helicity of particle $n$ to be positive, then the recursion relation is almost identical to the $\NN=4$ case:
\be\label{lessBCFW}
R^{(0)}_{n,k}=R^{(0)}_{n{-}1,k}+\sum_{k', i} R^{(0)}_{i,k'}(1,\cdots,i{-}1, I_i)\,[a,b,c,d,e]_{\NN}\, R^{(0)}_{n{+}2{-}i,k{-}1{-}k'} (-I_i,i,\cdots,\hat{n}^+),
\ee
where the shifted twistors are the same as above, and the helicity of $I_i$ depends on $k'$ and $i$~\cite{unpublished}, 
and the general $\NN<4$ five-bracket is defined as,
\be\label{lessRinvariant}
[a,b,c,d,e]_{\NN}\equiv \frac{\delta^{\NN}(\eta_a \l b c d e\r+\textrm{cyclic})}{\l a b c d\r\l b c d e\r\l c d e a\r\l d e a b\r\l e a b c\r}.
\ee

To see the soft theorem at work, note that although the R-invariant behaves like $\delta^{\NN{-}2}$ in the soft limit, the two sub-amplitudes will provide the additional powers of $\delta$. This is because, unlike $\NN=4$ amplitudes in momentum-twistor space, $\NN<4$ amplitudes carry non-zero weights for negative-helicity particles, which is the case for one of the $I_i$'s in the sub-amplitudes. Since $I_i\equiv \delta I'_i$, we have
\be
R^{(0)}_{i,k'}(1,\cdots,i{-}1, I_i) R^{(0)}_{n{+}2{-}i,k{-}1{-}k'} (-I_i,i,\cdots,\hat{n}^+)\sim\mathcal{O}(\delta^{4{-}\NN}), 
\ee
thus rendering these factorization terms again vanishing as $\delta^2$. 

At loop-level, integrands in $\NN<4$ SYM can also be obtained from \textit{e.g.} CSW diagrams~\cite{CSW,Bedford:2004nh}. For $\NN=4$ SYM, McLaughlin and one of the authors~\cite{HeTristan} proved that the integrand obtained from CSW diagrams are identical to the one from BCFW recursion relations above (see section~\ref{sec:F3} for generalization to $F^3$ amplitude). Given the similarity of the structures of integrands in $\NN=4$ and $\NN<4$, we conjecture that the soft theorem again holds already at the integrand level.

As an example which provides strong evidence for the conjecture, we now study one-loop amplitudes explicitly. The integrand for $\NN<4$ SYM amplitudes at one-loop can be written in terms of the one in $\NN=4$ and a part with $\NN=1$ chiral multiplets,  and it is sufficient to look at the soft behavior of the latter. A compact formula for the $\NN=1$ chiral part of the integrand has been written in momentum-twistor space using CSW diagrams~\cite{unpublished}: with $a, b$ the negative-helicity particles, the $\NN=1$ chiral part of the integrand, $R^{(1) {\rm chiral}}_{n,2}$, is given by 
\be\label{chiral}
R^{(1), \rm{chiral}}_{n,2}{-}R^{(1), \rm{chiral}}_{n{-}1,2}=-\frac{\l a \hat B\r\l b \hat B\r}{\l A B\r^2 \l AB 1 n{-}1\r\l AB n{-}1 n\r\l AB 1 n\r}\sum_{a<i\leq b}\frac{\l a I_i\r\l b I_i\r}{ \l AB 1 i{-}1\r\l AB i{-}1 i\r\l AB 1 i\r },
\ee 
where the hallmark of an $\NN=1$ chiral integrand is the appearance of the prefactor $1/\l A B\r^2=1/\l A B I\r^2$. 

The soft behavior of $R^{(1), \rm{chiral}}_{n,2}$ is given by $R^{(1), \rm{chiral}}_{n{-}1,2}$, plus the one of the R.H.S. of eq.~(\ref{chiral}). Recall that $I_i=\delta I'_i$ and $\hat{B}=\delta \hat{B}'$,  we see that the soft behavior is identical to the  $\NN=4$ case in eq.~(\ref{kermit}):
\ba
 &&-\frac{\l a \hat B\r\l b \hat B\r \l a I_i\r\l b I_i\r}{\l A B\r^2 \l AB 1 n{-}1\r\l AB n{-}1 n\r\l AB 1 n\r \l AB 1 i{-}1\r\l AB i{-}1 i\r\l AB 1 i\r }\nl
 =&&\frac{\delta^2}{\alpha\beta}\times  \frac{\l a \hat B'\r\l b \hat B'\r \l a I'_i\r\l b I'_i\r}{\l A B\r^2 \l AB 1 n{-}1\r^3 \l AB 1 i{-}1\r\l AB i{-}1 i\r\l AB 1 i\r}\,,
\ea
thus the soft theorem holds for one-loop MHV integrand in $\NN<4$ SYM. In addition, to obtain the $\NN=1$ chiral part for non-MHV amplitudes, one only needs to dress the above formula with two tree sub-amplitudes, so we conclude that the soft theorem, eq.~(\ref{integrandsoft}), holds for all one-loop amplitudes in $\NN<4$ SYM. 

Note that although the soft theorem is quite transparent using the BCFW-like recursion (when we shift the soft particle), it can be very non-trivial to see in terms of other representations of the same integrand, such as the local form based on leading singularities~\cite{ArkaniHamed:2010gh}.  For example, in that representation, the sub-leading terms cancel between different terms in a non-trivial way even for the one-loop integrand. 

More importantly, the soft theorem is generally not manifest at the integrand level for other representations, such as the form in \cite{Cachazo:2008vp} and \cite{Carrasco:2011mn}) for one-loop five-point amplitude in $\mathcal{N}=4$ SYM, which is given by scalar boxes and pentagon related to eq.~(\ref{kermit}) by integral reduction. The soft theorem is expected to hold only when we perform the integrals after the soft expansion. 

We have not discussed loop integrands in pure Yang-Mills theory, $\NN=0$, because it is not clear to us how to write down a four-dimensional integrand that manifest the soft theorem. It is also unclear how to apply our argument to cases where the definition of an integrand may be ambiguous, \textit{e.g.} non-planar theories such as gravity. In section~\ref{Allplus}, we will discuss the soft theorem with the integrals performed, for the case of all-plus amplitudes in both YM theory and gravity. 

\section{Soft theorems for finite loop amplitudes}\label{Allplus}
We now consider cases where the integrand does not manifestly satisfy the soft theorem, and thus integration is required. As discussed in the introduction, the loop-level soft theorem can be formulated in two distinct prescriptions: (1) taking $\epsilon\rightarrow0$ before expanding in the soft parameter $\delta$, or (2) first expand the integrand in the soft parameter $\delta$, and then perform the integration with the regularization. For general integrands the two limits do not commute as was pointed out in \cite{FreddyEllis}. That this is the case can be simply understood from the fact that soft expansion of the integrand assumes that the loop momentum is hard compared to the soft external momenta. This assumption becomes untenable in the region where the loop-momentum itself is soft, which is precisely the region to be regulated by $\epsilon$. Thus for the soft behavior of a loop amplitude, it is more convenient to implement procedure $(1)$. We refer to~\cite{FreddyEllis} for a detailed discussion on this issue.

On the other hand, whether or not the soft theorem is indeed un-renormalized in the context of prescription (2), is an interesting question on its own right and can provide non-trivial constraint on the integrand. Indeed as we've seen from previous discussions, the planar-integrand of $\mathcal{N}\leq4$ SYM manifestly respects the un-renormalized soft theorems prior to integration. In this section, we would like to take preliminary steps in extending the discussion to pure YM and gravity amplitudes. We will consider one-loop amplitudes where the soft-behaviour is non-renormalized in both prescriptions, the all-plus YM and gravity amplitudes. We will show that agreement of the two approaches is precisely due to the fact that the relevant integrands enjoy the property that the two limits commute.

We will also discuss why such commutative property no longer holds for the single-minus amplitude. The violation of tree-level soft theorems for single minus amplitude can also be understood from the classical symmetries and its violation at loop level.

\subsection{All-plus Yang-Mills amplitude}
The $D$-dimensional all-plus integrand can be obtained straightforwardly from the $\mathcal{N}=4$ SYM integrand by simply multiplying it by extra powers of the regulator mass $(\mu^2)^2$~\cite{ZviYMAllplus}. Naively, since we have already shown that the planar integrand vanishes for the sub-leading term in the kinematic configuration of eq.(\ref{integrandsoft}), multiplying by an overall factor would not change this result. However, as one convert the momentum twistor integrand into momentum space, the non-uniqueness of the identification of $\ell$ obscures this property and integration is necessary to show the vanishing of the sub-leading  terms.

Let us first consider the one-loop five-point all-plus amplitude. The $D$-dimensional integrand is given as~\cite{ZviYMAllplus}:
\eq\label{AllPlusInt}
A_{5}^{+,+,+,+,+}=\frac{2}{\prod_{i=1}^5 \la ii+1\ra}\left(-\frac{1}{2}\left[\frac{\mu^4 s_{12}s_{23}}{d_1d_2d_3d_5}+cyclic\right]+\frac{4i\mu^6\epsilon(1234)}{d_1d_2d_3d_4d_5}\right)\,,
\eqe
where $d_i=\ell_i^2$ and $\ell_i=\ell+\sum_{j=1}^i k_i$, and thus $\ell$ is positioned between $5$ and $1$. In the soft-limit, the numerators of the above integrand behaves as:
\eqa
s_{12}s_{23}&=&s_{1'2}s_{23}+ \delta s_{23}s_{2 p_1},\quad s_{23}s_{34}=s_{23}s_{34'}+ \delta s_{23}s_{3 p_4},\nonumber\\
 s_{34}s_{45}&=&\delta s_{34'} s_{54'}+\mathcal{O}(\delta^2), \quad s_{45}s_{51}=\mathcal{O}(\delta^2),\quad s_{51}s_{12}=\delta  s_{1'2}s_{51'}+\mathcal{O}(\delta^2)\nonumber\\
 \epsilon(1,2,3,4)&=&\delta\epsilon(p_1,k_2,k_3,k_{4'})+\delta\epsilon(k_{1'},k_2,k_3,p_{4})+\mathcal{O}(\delta^2)
\eqae
where $s_{i p_j} = (k_i + p_j)^2$ and we have used the notation:
\eqa
k_1&=&k'_1+\delta p_1,\quad k'_1=-|1\ra\sum_{i=2,3}\frac{\la 4 i\ra}{\la 41\ra}[i|, \quad p_1=-|1\ra\frac{\la 4 5\ra}{\la 41\ra}[5|\nonumber\\
k_4&=&k'_4+\delta p_4,\quad k'_4=-|4\ra\sum_{i=2,3}\frac{\la 1 i\ra}{\la 14\ra}[i|, \quad p_4=-|4\ra\frac{\la 1 5\ra}{\la 14\ra}[5|\,.
\eqae
Note that $k_{1'}+k_2+k_3+k_{4'}=0$. Since the Parke-Taylor prefactor behaves as $1/\delta^2$, the leading soft contribution comes from, the first two terms in the square bracket in eq.(\ref{AllPlusInt}), which indeed is $S^{(0)}A_{4}^{+,+,+,+,+}$ at the integrand level. For the sub-leading  term, again only for the first two terms in the square bracket does one need to soft expand the integrand. Note that since the integrand integrates to a constant, there are no sub-leading  contribution if one follows prescription (1). On the other hand since 
\eq
I_{m}[\mu^{2r}]=-\epsilon(1-\epsilon)\cdots(r-1-\epsilon)(4\pi)^{r}I_{m}^{D=4+2r-2\epsilon}\,,
\eqe
the fact that the pre-expanded integral is a constant implies that the $I_{m}^{D=4+2r-2\epsilon}$ in the above is logarithmic divergent. The soft expansion then introduces an additional propagator which would render $I_{m}^{D=4+2r-2\epsilon}$ finite, leading to a vanishing result as well.  Thus to order $\epsilon$, the two prescriptions agree and the soft-theorem is non-renormalized in both cases.

The same analysis applies to general $n$. As $\mathcal{N}=4$ SYM contains no triangles or bubbles, dimension shifting formula tells us that the all-plus integrand can be simply expressed in terms of scalar boxes and pentagons multiplied by $(\mu^2)^2$. The sub-leading  soft expansion of these integrals vanishes, in agreement with the expansion of the integrated results. Note that if the integrand includes scalar triangle and bubbles, $I_{3}[\mu^{2r}]$ and $I_{2}[\mu^{2r}]$, the two limits may no longer commute. This is due to the fact that the soft expansion can introduce scale free integrals which strictly integrate to zero in dimension regularization, but are of order $\delta$ if one expands the integrated result. A trivial example would be the following bubble integral:
$$\includegraphics[scale=0.5]{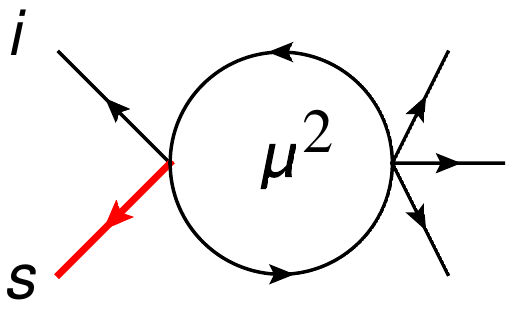}$$ 
which integrates to $k^2_{si}$, and thus becomes of order $\delta$ in prescription $(1)$, while in prescription (2) it integrates to $\delta\times 0$, since in the soft limit, the integrand becomes a massless bubble integral. Similarly for $I_{3}[\mu^{4}]$, if the soft leg is on a massless corner the soft expansion is of order $\delta$ in prescription $(1)$ while vanishes in prescription (2). The possible disagreement of soft theorems between prescription (1) and (2) for the single minus amplitude can be traced to the presence of these integrals in the final answer. Indeed already at four-points $A_4(-,+,+,+)$ contains the bubble integrals mentioned above~\cite{BernMassive}.

\subsection{All-plus Gravity amplitude}
We now consider all-plus gravity amplitudes. The integrand is given by dimension-shifting formulas from one-loop MHV amplitude in $\mathcal{N}=8$ supergravity~\cite{BernAllPlus}. Note that due to higher powers of $\mu^2$, the fact that the two limits commute is rather non-trivial. Consider 
\eq
M_5=\beta^{123(45)} I^{123}[(\mu^2)^4] +\gamma^{12345}I^{12345}[(\mu^{2})^{10}] +{\rm Perm}
\eqe
where 
\eq
\beta^{123}=-\frac{[12]^2[23]^2[45]}{\la 14\ra \la 15\ra \la 34\ra \la 35\ra \la 45\ra},\quad\gamma^{12345}=-2\frac{[12][23][34][45][51]}{\la 12\ra \la 23\ra \la 34\ra \la 45\ra \la 51\ra }
\eqe
one sums over 30 inequivalent box integrals and 12 pentagons. First let's consider to which order in $\delta$ one should expand the integrals in the above representation. First of all for the pentagon, since the prefactor begins at order $\delta^{-2}$, for the first sub-leading  behavior of the integrand, we do not need to expand the pentagon integrand.  For the box-integrals, there are three distinct types to consider in the soft-limit: (I) if the soft leg is on the massive corner, there are 12 such diagrams. (II) if the soft leg is on the massless corner adjacent to the massive corner, there are again $12$ such diagrams. (III) the soft leg is diagonal to the massive corner, which consists of 6 diagrams. The coefficient for the last case (III) behaves as $\mathcal{O}(\delta^0)$ in the soft limit and thus will not participate in the discussion. The pre factor for case (II) behaves as $\mathcal{O}(\delta^2)$ and thus there is no need to expand the integrand. Finally, case (I) is of order $\frac{1}{\delta^3}$, and thus we need the result of the integral expanded to order $\delta$. Denoting the integrand by its three massless legs $I_4(i,j,k)$:
\eq
I_4(i,j,k) =\quad 420\times\vcenter{\hbox{\includegraphics[scale=0.45]{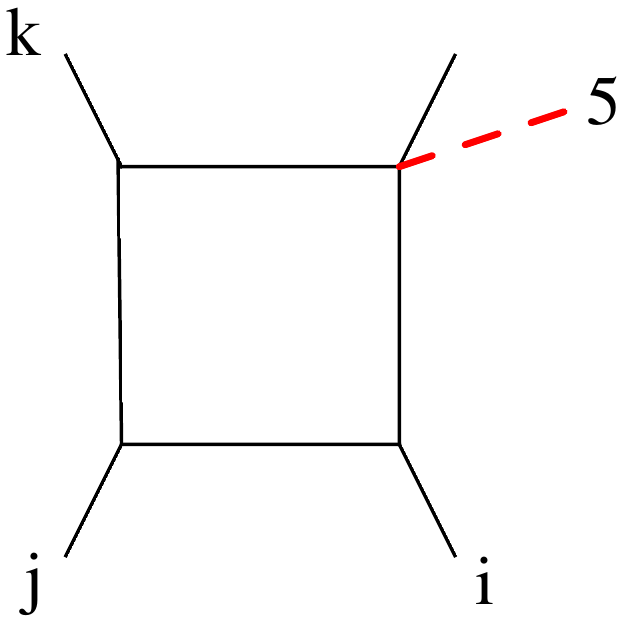}}}\,.
\eqe
We list the order $\mathcal{O}(\delta)$ contribution in the following table
$$\begin{array}{|c|c|}\hline \;  & \mathcal{O}(\delta^1)  \\\hline I_4(1',2,3) & {-}2u(p_1\cdot k_4'){-}(4s{+}t)(p_1\cdot k_2)  \\\hline  I_4(3,1',2) &  -(6s{+}3u)(p_1\cdot k_2){-}(6u{+}3s)(p_1\cdot k_3)   \\\hline I_4(3,4',1') & \begin{array}{c}{-}(4s{+}t)(p_4\cdot k_3){-}(5s+12t)(p_4\cdot k_1'){-}(7s+14t)(p_1\cdot k_4') \\ {-}2[u(p_5\cdot k_3){+}(2s{+}4t)(p_5\cdot k_4'){+}(s+3t)(p_5\cdot k_1')]\end{array}\\\hline I_4(1',3,4') & {-}(6u{+}3s)(p_1\cdot k_3){+}2t(p_1\cdot k_4'){+}2t(p_4\cdot k_1'){-}(3u+6s)(p_4\cdot k_3)   \\\hline \end{array}$$
while all others are related by symmetry. It is straightforward to check that the above result is the same as $\mathcal{O}(\delta^1)$ of:
\eq
I_4(1,2,3)=-\frac{2s_{12}^2+2s_{23}^2+2(K^2)^2+s_{12}s_{23}+2s_{12}K^2+2s_{23}K^2}{2}\,,
\eqe
where $K$ is the momenta on the massive leg. Thus we see for the sub-leading  soft contribution, the two prescriptions again commut and the soft theorem is unrenormalized in both descriptions. 

The above analysis should come as no surprise given the fact that the integrals involved remain finite, whether or not the soft expansion is done before or after the integration and thus the limits should commute.  Again for bubble and triangle integrals, the two-limits no longer commutes, thus the fact that the soft theorem for the all-plus gravity amplitude agrees in both prescription can be associated with the fact that the dimension shifting formula allows only box and pentagon integrals in the representation.  

\subsection{Conformal anomaly and the integrated soft theorems}\label{conformal}
An alternative way to understand why the integrated soft theorems for all plus amplitude is not corrected, while the single-minus are is via symmetries.  Indeed it was demonstrated in~\cite{Andrew} that given the leading soft function, with suitable assumptions the sub-leading soft operator is determined by the conformal symmetry of tree-level amplitude. Thus the tree-level soft-functions can be viewed as the homogenous solutions to the differential equation implied by the symmetry constraints. From this point of view, the loop-level corrections can be attributed to the fact that this symmetry becomes anomalous at loop level. In particular, since the all-plus amplitude is generated by the self-dual sector of Yang-Mills theory, it is protected and conformal symmetry is preserved implying that the soft function is not corrected. For single-minus amplitude, this is no longer the case and potential correction terms arrise, as verified in~\cite{HHW, ZviScott}.

To see this, note that conformal symmetry of the $(n{+}1)$-point amplitude implies\footnote{Unlike other sections, here we put a superscript ``tree" on $S^{\rm tree (i)}_{\rm YM}$ to emphasize they are tree-level results, and we will consider corresponding loop corrections.} 
\eq\label{Target}
( \mathfrak{K}_0+\frac{1}{\delta} \mathfrak{K}_s)(\frac{1}{\delta^2}S_{\rm YM}^{\rm tree (0)}A_n+\frac{1}{\delta}S^{\rm tree (1)}_{\rm YM}A_n)=0
\eqe
where we've separated the conformal boost generator into
\eq\label{X1}
 \mathfrak{K}_0=\sum_{i=1}^n \frac{\partial}{\partial \lambda_i} \frac{\partial}{\partial \tilde\lambda_i},\quad  \mathfrak{K}_s= \frac{\partial}{\partial \lambda_s} \frac{\partial}{\partial \tilde\lambda_s}\,,
\eqe
where we've suppressed the Lorentz indices $\alpha, {\dot\alpha}$.
Now starting with $S^{(0)}=\frac{\la n1\ra}{\la ns\ra\la s1\ra}$, at order $\mathcal{O}(\delta^{-3})$ eq.(\ref{Target}) is trivially satisfied, while at $\mathcal{O}(\delta^{-2})$ we have the following constraint:
\eq
 \mathfrak{K}_0S_{\rm YM}^{\rm tree (0)}A_n+\mathfrak{K}_sS^{\rm tree (1)}_{\rm YM}A_n=-\left(\frac{\lambda_n}{\la ns\ra^2}\frac{\partial}{\partial \tilde{\lambda}_n}+\frac{\lambda_1}{\la 1s\ra^2}\frac{\partial}{\partial \tilde{\lambda}_1}\right)A_n+(\mathfrak{K}_sS^{\rm tree (1)}_{\rm YM})A_n=0
\eqe
One can check that the tree-level soft function $S^{\rm tree (1)}_{\rm YM}$ is the homogenous solution to the above conformal boost equation. The same analysis applies to the super soft-functions as we show in appendix~\ref{sec:SuperSoft}.

A consequence of this analysis is that if conformal symmetry becomes anomalous, as one expects at loop level, then the soft function has to be modified. Let's consider the conformal boost equations in the presence of anomalies:
\eq\label{Anom}
(\mathfrak{K}_0+\frac{1}{\delta}\mathfrak{K}_s)A_{n+1}(\{\lambda_i,\tilde{\lambda}_i\}, \delta\lambda_s,\tilde{\lambda}_s)=\sum_{i} a^{(i)}_{n+1}\delta^i
\eqe
where $a_i$'s are the conformal anomaly expanded in the soft parameter. We begin with the following ansatz for the soft expansion of $A_{n+1}$, 
\eq
\sum_{i=0}^1\frac{1}{\delta^{i{+}1}}S_{\rm YM}^{\rm tree (i)}A_{n}+\Delta^{(i)}+\mathcal{O}(\delta^0)\,.
\eqe
From eq.(\ref{Anom}), we have the following constraints on the unknown function $\Delta^{(i)}$:
\eqa\label{AnomC}
\nonumber \mathcal{O}(\delta^{-3})&&\quad \mathfrak{K}_s(S_{\rm YM}^{\rm tree (0)}A_{n}+\Delta^{(0)}) =a^{(-3)}_{n+1},\\
 \mathcal{O}(\delta^{-2})&&\quad \mathfrak{K}_0(S_{\rm YM}^{\rm tree (0)}A_{n}+\Delta^{(0)})+\mathfrak{K}_s(S_{\rm YM}^{\rm tree (1)}A_{n}+\Delta^{(1)})=a^{(-2)}_{n+1}\,.
\eqae
Now as the all-plus amplitude is associated with the self-dual sector of YM theory which is exact, this implies that its amplitude is conformally invariant. Thus we expect no correction to the soft functions, {\it i.e.} $\Delta^{(i)}=0$. For single minus amplitude, this is no-longer true and potential correction terms may arise. It is straight forward to verify that in the soft limit, if the soft leg is minus helicity the anomaly is finite, and thus eq.(\ref{AnomC}) reduces to zero on the RHS, leading to the conclusion that one only has the tree-level soft theorem. For the negative helicity leg the anomaly begins at $\delta^{-2}$. The absence of $a^{(-3)}_{n+1}$ infers $\Delta^{(0)}=0$, and thus $\Delta^{(1)}$ must satisfy
\eq\label{AnomEq}
\mathfrak{K}_s(\Delta^{(1)})=a^{(-2)}_{n+1}-a^{(0)}_{n}
\eqe
The explicit correction term for the single minus amplitude is given in~\cite{HHW}:
\eqa\label{SingleMinusD}
\Delta^{(1)}&=&-\frac{\la n1\ra^4}{\prod_{i=1}^n\langle ii+1\rangle}\frac{\la n-1s\ra[nn+1]}{\la n-1n\ra\la ns\ra^2}
\eqae
We have explicitly verified that the above expression indeed satisfies eq.(\ref{AnomEq}).

\section{Soft theorems for higher-derivative interactions}\label{higher-dim}
In this section, we would like to consider to which extent the soft theorem is universal for tree-level scattering amplitudes of Yang-Mills and gravity theories coupled matter, or effective field theories with higher-dimensional operators. For the later, it can be viewed as posing the same question for tree-level string-theory amplitudes in the $\alpha'$ expansion. Recall that from Low's work the soft gluon/graviton behavior of perturbative scattering amplitudes is determined by the three-point interaction of the theory and gauge invariance, thus one expects that only higher-dimensional operators that modify the three-point interaction is relevant to the discussion. While such interactions are generically suppressed in the soft limit by extra power of soft invariants, this does not rule out the possibility of modification in the sub-leading behaviors.

Here we will only consider higher-dimensional operators that involve massless fields. For massive fields, the soft behavior is non-trivial at orders beyond that under discussion for soft theorems. With that in mind we will consider amplitudes arising from $F^3$, $R^3$ and $R^2\phi$, where the scalar field is a massless dilaton.  
\subsection{Amplitudes from $F^3$}~\label{sec:F3}
We first consider $F^3$ operator, whose general multiplicity amplitudes were studied in~\cite{F3R3Lance}. We will consider amplitudes that are generated by the self-dual contribution from a single $F^3$. Here, by self-dual we are referring to the part of the $F^3$ that produces an all-minus three-point amplitude. 
\subsubsection*{CSW representation of $F^3$ Amplitudes }
Using a CSW representation~\cite{CSW}, the $n$-point $k$-minus helicity amplitude is given by a single $F^3_{\rm SD}$ vertex connected with $(k{-}3)$- YM MHV vertices. Thus there are two types of vertices in the CSW rule: (1) a white vertex representing a $F^3_{\rm SD}$ vertex, with its associated ``MHV" building block given by:
\eq
\vcenter{\hbox{\includegraphics[scale=0.7]{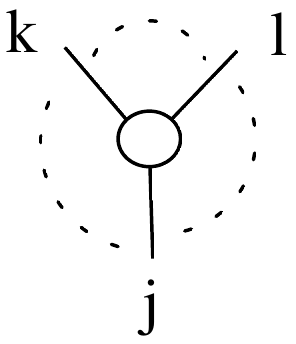}}}:\quad \frac{\la jk\ra^2 \la kl\ra^2\la lj\ra^2}{\prod_{i=1}^n\la ii+1\ra}
\eqe 
where the lines $j,k,l$ are the negative helicity legs, while the dots represent positive helicity legs. (2) a black vertex representing the usual YM MHV vertices:
\eq
\vcenter{\hbox{\includegraphics[scale=0.7]{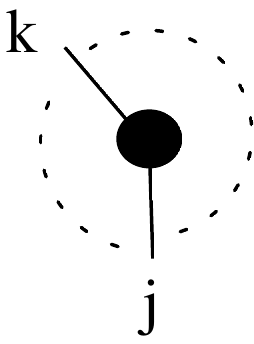}}}:\quad \frac{\la jk\ra^4}{\prod_{i=1}^n\la ii+1\ra}\,.
\eqe
Here we will consider diagrams with only one white vertex. For example the NMHV amplitude consists of two diagrams (here, N$^k$MHV refers to $k+3$ minus helicity legs):
\eq
(a):\;\;\vcenter{\hbox{\includegraphics[scale=0.5]{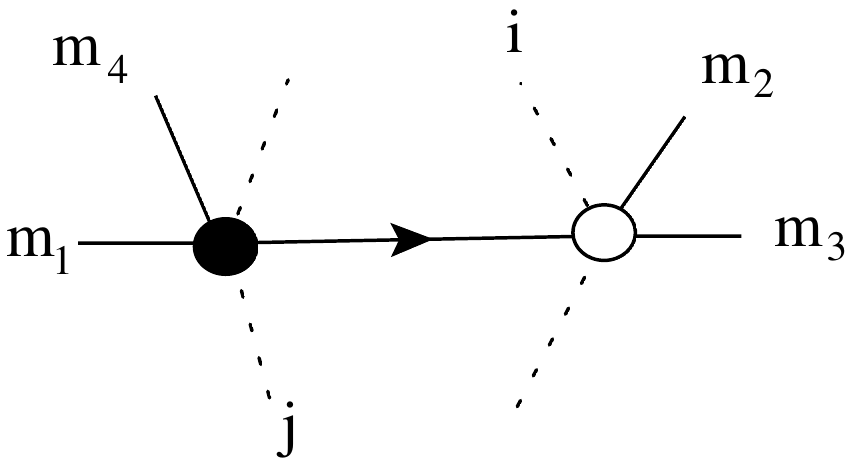}}},\quad\quad (b):\;\; \vcenter{\hbox{\includegraphics[scale=0.5]{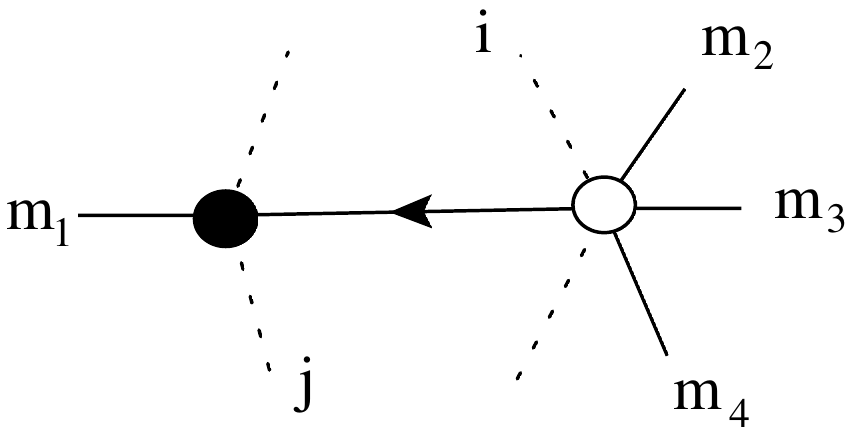}}}
\eqe
where the arrows on the propagator indicate to which vertex the negative helicity is associated. The dotted lines simply represents the legs that are adjacent to the propagator, and can be one of the minus legs. It is convenient to pull out an overall Parke-Taylor factor, so that the contributions from the above two diagrams are given by:
\eqa\label{PreMHV}
\nonumber (a):&&\;\;\frac{1}{\prod_{l=1}^n\la ll{+}1\ra}\left(\frac{\la m_1m_4\ra^4}{\la i{-}1 P\ra\la Pj\ra}\frac{\la i{-}1i\ra\la j{-}1j\ra}{P^2}\frac{\la m_2m_3\ra^2\la m_3P\ra^2\la Pm_2\ra^2}{\la P i\ra\la j{-}1P\ra}\right)\\
\nonumber (b):&&\;\;\frac{1}{\prod_{l=1}^n\la ll{+}1\ra}\left(\frac{\la m_1P\ra^4}{\la i{-}1P\ra\la Pj\ra}\frac{\la i{-}1i\ra\la j{-}1j\ra}{P^2}\frac{\la m_2m_3\ra^2\la m_3m_4\ra^2\la m_4m_2\ra^2}{\la P i\ra\la j{-}1P\ra}\right)\,.
\eqae
where $\la P|=P|\mu]$ for some reference spinor $|\mu]$. 

\subsubsection*{$F^3$ amplitudes in momentum twistor space and recursions}
To facilitate the analysis, we will now convert the expressions into momentum twistor space.  This will allow us to reveal the fact that amplitudes of $F^3$ operator with at least one plus helicity leg respects a BCFW recursion. The momentum twistor space CSW prescription for on-shell spinors are as follows. Consider a propagator connecting two vertices defined by two regions (i,j). In momentum twistor space, they are given by:
\eq\label{StarDef1}
\vcenter{\hbox{\includegraphics[scale=0.5]{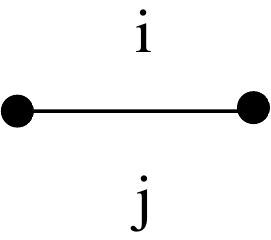}}}:\quad \langle a P\rangle\equiv \frac{\langle a [ i\ra \la i{-}1] jj{-}1*\ra}{\la ii{-}1\ra\la jj{-}1\ra}=-\frac{\langle a [ j\ra \la j{-}1]  i i{-}1*\ra}{\la ii{-}1\ra\la jj{-}1\ra}
\eqe
where the equality holds due to the fact that the reference twistor $Z_*=(0,\mu,0)$. If two propagators are connected to the same vertex and adjacent, one then has: 
\eqa\label{StarDef2}
\vcenter{\hbox{\includegraphics[scale=0.5]{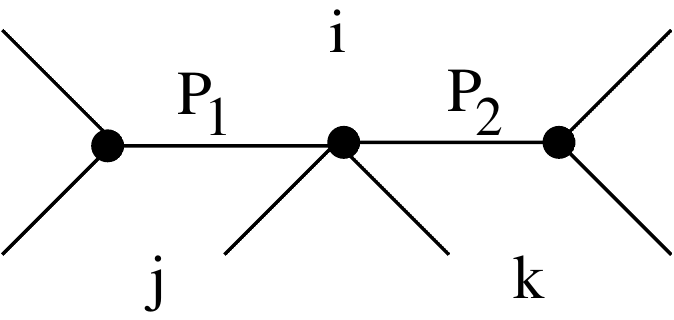}}}\quad\la P_1P_2\ra&=&\frac{\la i i{-}1\,(* jj{-}1)\cap(*kk{-}1)\ra}{\la ii{-}1\ra\la jj{-}1\ra \la kk{-}1\ra}\equiv-\frac{\la * jj{-}1[i\ra\la i{-}1] kk-1*\ra}{\la ii{-}1\ra\la jj{-}1\ra \la kk{-}1\ra}\nonumber\\
&=&\frac{\la * kk{-}1\widehat{i{-}1}\ra}{\la ii{-}1\ra\la jj{-}1\ra \la kk{-}1\ra}
\eqae
where in the final line $\widehat{i{-}1}\equiv (ii{-}1)\cap(*jj{-}1)$. These will be the fundamental identifications used throughout. 

Using these identities, we find that the amplitudes in eq.(\ref{PreMHV}) can be rewritten in the following succinct form:
\eqa\label{PreMHV}
\nonumber (a):&&\;\;\frac{1}{\prod_{l=1}^n\la ll{+}1\ra}  \la m_1m_4\ra^4 \;  [ii{-}1jj{-}1*] \; \la m_2m_3\ra^2\la m_3 \widehat{i{-}1}\ra^2\la \widehat{i{-}1} m_2\ra^2\\
\nonumber (b):&&\;\;\frac{1}{\prod_{l=1}^n\la ll{+}1\ra}\la m_1\widehat{i{-}1}\ra^4\; [ii{-}1jj{-}1*] \; \la m_2m_3\ra^2\la m_3m_4\ra^2\la m_4m_2\ra^2\,,
\eqae
where recall from eq.~(\ref{lessRinvariant}) $[ii{-}1jj{-}1*] $ is defined as (here $\NN=0$):
\eq
[abcde]\equiv\frac{1}{\la abcd\ra\la bcde\ra\la cdea\ra\la deab\ra\la eabc\ra}\,. 
\eqe
Thus for any CSW diagram, one simply replace each propagator by a factor of $[* i i-1 j j-1]$, while each black or white vertex is dressed with:
\eq
\vcenter{\hbox{\includegraphics[scale=0.5]{WhiteVertex}}}:\quad\quad \la jk\ra^2 \la kl\ra^2\la lj\ra^2,\quad\vcenter{\hbox{\includegraphics[scale=0.5]{BlackVertex}}}:\quad\quad \la jk\ra^4\,.
\eqe

Equipped with the momentum twistor space representation, we will now show that if there is at least one plus helicity leg, the result from CSW construction satisfies the BCFW recursion relation similar to that in Yang-Mills theory (we use $R$ to represent amplitudes with an overall $(\prod_i\la i i+1\ra)^{-1}$ stripped off):
\eqa
\label{TreeBCFWRecur}
R^{F^3}_{k,\,n}&=&R^{F^3}_{k,\,n{-}1}+\sum_j [n{-}1,n,1,j{-}1,j] R^{F^3}_{k',\,j}(1,\cdots, I_j) R^{F^2}_{k{-}1{-}k',\,n{+}2{-}j}(-I_j,\cdots, \hat{n}_j) \nonumber\\
&+& (F^3\leftrightarrow F^2)~,
\eqae
where $2<j<n$, $I_j=(j-1 j)\cap(n{-}1,n,1)$, $\hat{n}_j=(n{-}1,n)\cap (1,j{-}1,j)$, and similar to above, we have assumed leg $n$ to be positive-helicity. Note that in momentum space this corresponds to the $[n-1 n\rangle$ shift, for which we have explicitly checked that up to six points, the amplitudes listed in~\cite{F3R3Lance} indeed vanish at $z\rightarrow \infty$.

The proof proceeds exactly as that of $\mathcal{N}=4$ SYM~\cite{HeTristan}, namely by judiciously choosing the reference twistor, one can show that the difference between the $n{+}1$- and $n$-point CSW representation, $R^{F^3}_{k,\,n}-R^{F^3}_{k,\,n-1}$, is given by the last term in eq.(\ref{TreeBCFWRecur}). First note that as the twistor $Z_n$ is a positive helicity leg, it generically does not appear in the two expressions, and hence most of the terms cancel immediately. Let us first consider NMHV tree, where the mismatch is given simply by
\eqa
\nonumber R^{F^3}_{k,\,n}-R^{F^3}_{k,\,n-1}&=&\left(\sum_j [*,n{-}1,n,j-1,j]\bar{X}(n{-}1, n,j)+\sum_{j}[*,n,1,j-1,j]\bar{X}(n,1,j)\right.\\
-&&\left.\sum_{j}[*,n{-}1,1,j-1,j]\bar{X}(n{-}1,1,j)\right)\,,\label{TreeSum}
\eqae
where $\bar{X}$ simply denote the vertex factors for each diagram. Now if we take $Z_*=Z_1$, the last two terms vanish. To be more precise, while the denominator of $[*,n{-}1,1,j-1,j]$ contains three zeroes, the factor $\bar{X}$ contains four factors of $\la a_iP\ra=\la a_i [j\ra\la j{-}1] n1*\ra$ which vanishes as $*=1$. Thus the CSW representation for the NMHV tree-level amplitude is simply given as 
\eqa\label{NMHVtreeCSW}
R^{F^3}_{k,\,n}&=&R^{F^3}_{k,\,n-1}+\sum_{j}[n{-}1, n, 1j-1,j]\bar{X}(n{-}1,n,j)\,.
\eqae
Note that the factors in $\bar{X}$ which involves the propagator leg $|P\ra$ is evaluated at $(j-1j)\cap(n{-}1, n,1)$, {\it i.e.} it is given by $\hat{I}_i$. Furthermore, since leg $n$ has positive helicity, it does not appear explicitly in the above representation and we are free to make the identification for $\hat{n}_j$.

For general N$^{\rm k}$MHV amplitude, the proof of equivalence again simply follows that of $\mathcal{N}=4$ SYM given in \cite{HeTristan}. The classification of all CSW digram is given by a collection of $2k$ set of region momenta, separated into $k$ non-crossing pairs. The difference $R^{F^3}_{k,\,n}-R^{F^3}_{k,\,n-1}$ is given by CSW diagrams where one of the non-crossing pairs are $(2,i)$.The remaining pairs factorize. Distinct choices of $i$ can then be mapped into distinct helicity distributions in the BCFW recursion. Again the only difference between the $\mathcal{N}=4$ and the present case is the presence of the $\bar{X}$ factors arising from each vertex. 

\subsubsection*{Soft limits of $F^3$ amplitudes }
We now consider the soft limits of $F^3$ amplitudes. Note that the recursion formula derived from above assumes that there is at least one plus helicity leg, $n$. This is no longer valid for the all-minus amplitude that is also generated by $F^{3}_{SD}$. Fortunately, it is straight-forward to study the soft minus gluon limit in the CSW representation, since the only place where anti-holomorphic spinors appear in the CSW representation is in the propagators and $\la P|$. With generic reference spinor, the only singularities that appear are associated with the soft leg attached to a three-point vertex with another external leg. If the three-point vertex is an $F^3$, then one has:
\eqa
\nonumber \vcenter{\hbox{\includegraphics[scale=0.5]{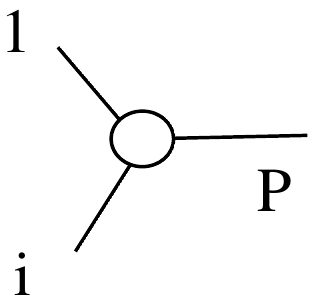}}}:&&\quad \frac{\la 1P\ra\la Pi\ra}{[i1]}\rightarrow  \frac{\la 1 i\ra[i\eta][\eta 1]\la 1i\ra}{ [i1]}
\eqae
which is finite for the soft leg 1. This is just a reflection of the fact that $F^3$ operator is higher dimensional and suppresses the soft divergence. If the three-point vertex is the usual MHV vertex, then the soft theorem simply follows from Low's analysis (or by expanding MHV diagrams to the subleading order).

Let us now consider the recursion in eq.(\ref{TreeBCFWRecur}), and take the positive helicity leg-$n$ to be soft. The discussion parallels that for YM tree amplitudes above: the BCFW shifted variables behave as $\hat{I}_j=\delta\,(j{-}1j)\cap(n{-}1,s,1)$, while all other variables remain unchanged.\footnote{Except for $\hat{n}_j=(n{-}1,n)\cap (1,j{-}1j)+\mathcal{O}(\delta)$. But $\hat{n}_j$ never explicitly appear in the expression.}  Let us first look at the factorization terms in eq.(\ref{TreeBCFWRecur}). The pre-factor $[n,1,2,j{-}1,j]$ behaves as $\delta^{-2}$:
\eq
[n{-}1,n, 1,j{-}1,j]=-\frac{1}{\delta^2\alpha \beta\la n{-}1\, 1\, s\, j{-}1\ra\la n{-}1\, 1\, s\, j \ra\la n{-}1 \, 1\, j{-}1\, j \ra^3}+\mathcal{O}(\delta^{-1})\,.
\eqe
On the other hand, $\hat{I}_j$ appears in the tree amplitude on both sides as $\la \hat{I}_j x\ra$ with degree 4 in $\la \hat{I}_j|$. Thus the overall result of the factorization terms is of degree $\mathcal{O}(\delta^2)$. Thus in the anti-holomorphic soft limit, we find 
\eq
R^{F^3}_{k,\,n}=R^{F^3}_{k,\,n-1}+\mathcal{O}(\delta^{2})\,,
\eqe
as dictated by the original tree-level Yang-Mills soft theorem.

\subsection{Higher-derivative gravitational  interactions and their soft limits}
From the previous discussion, we have seen both via heuristic arguments and explicit analysis that higher derivative operators do not modify soft theorems due to their suppression at small momenta. Extending the argument to gravity, one would reach the same conclusion as gravity operators are further suppressed. However, it is easy to see that this is not always true. Consider the tensoring of two $F^3$ scattering amplitudes via KLT relations~\cite{KLTref}. The explicit amplitude up to six-point was given in~\cite{F3R3Lance}. Take for example 
\eq\label{M5}
M(1^-,2^-,3^-,4^-,5^+)=is_{12}s_{34}A^{F^3}(1^-,2^-,3^-,4^-,5^+)A^{F^3}(2^-,1^-,4^-,3^-,5^+)+\mathcal{P}(2,3)
\eqe
It is straightforward to verify that 
\eq
M(1^-,2^-,3^-,4^-,5^+)\bigg|_{ \lambda_5\rightarrow \delta \lambda_5}=\sum_{i=0}^2 \frac{1}{\delta^{3-i}}S^{(i)}_{\rm G}(5)M_4+\mathcal{O}(\delta^0)\,,
\eqe
where $M_4=M_4(1^-,2^-,3^-,4^-)$. However, taking the anti-holomorphic soft limit on leg $1$, we find:
 \eq\label{Mismatch}
M(1^-,2^-,3^-,4^-,5^+)\bigg|_{ \tilde\lambda_1\rightarrow \delta \tilde\lambda_1}=\sum_{i=0}^2 \frac{1}{\delta^{3-i}}S^{(i)}_{\rm G}(1)M_4+\frac{1}{\delta}\Delta^{(2)}+\mathcal{O}(\delta^{0})\,,
\eqe
where, now, $M_4=M_4(2^-,3^-,4^-,5^+)$, and $\Delta^{(2)}$ is an unknown correction to $S^{(2)}_{\rm G}$ for now. The fact that $S^{(2)}_{\rm G}$ is violated can be traced back to the presence of a dilaton exchange induced by the higher-dimensional operator $\phi R^2$. Using string theory language the operator $F^3$ is of order $\alpha'$, and thus via KLT one obtains an amplitude that is of order ${\alpha'}^{2}$ in the effective field theory. This receives contribution form $R^3$, which is of order ${\alpha'}^{2}$, and two insertions of $\phi R^2$, each of order $\alpha'$. Let's consider the exchange of a dilaton between a $\phi R^2$ vertex and a tree-diagram associated with a single $\phi R^2$ operator. In the mostly minus amplitude, the two gravitons on the $\phi R^2$ vertex must be of negative helicity, and the contribution is proportional to:
\eq
\frac{\langle 12\rangle^3}{[12]}\times M_n(\tilde{\phi})\,,
\eqe 
where $M_n(\tilde{\phi})$ is a tree-level diagram with the dilaton leg off-shell. 
As one can see taking either leg to be soft, one finds a $\frac{1}{\delta}$ contribution proportional to the tree-level amplitude generated by $\phi R^2$. The latter can be easily obtained by KLT tensoring $F^3$ amplitude with usual YM $F^2$ amplitude. Indeed the modification for $S^{(2)}_{\rm G}$ is precisely given by:
\eq
\Delta^{(2)}=\sum_{j}-2\frac{\langle 1j\rangle^3}{[1j]}M_n(\phi, i^-_{1}, i^-_{2},\cdots, i^-_{n-2}, n^+)\,,
\eqe
where $j$ runs over all remaining minus helicity legs, and $(i^-_{1}, \cdots, i^-_{n-2})\neq j$. With this modification we indeed reproduce the correct $\delta^{-1}$ term in eq.(\ref{Mismatch})\footnote{We will find the same conclusion in section \ref{BCFWstring} for bosonic closed-string amplitudes via BCFW recursion relations.}. Note that this also explains why the plus-helicity soft limit of the amplitude in eq.(\ref{M5}) does not require modification: for the presence of $\phi R^2$ to appear in the positive helicity soft channel, there must be at least two positive heliclity legs. Such corrections to the sub-leading term is very similar to the corrections present in the single-minus amplitude of QCD~\cite{HHW}, where the correction term is proportional to a lower-point amplitude with one of the states replaced due to the presence of a new effective vertex.

While the above operators can be ruled out at tree level for supersymmetric theories, such operators can still be generated via anomalies at loop level in supergravity theories. Indeed the $U(1)$ anomaly in $\mathcal{N}=4$ supergravity is known to generate a term in the effective action that is of the form $(R^{+})^2\bar{t}$~\cite{RaduAnomaly}, where $R^+$ is the anti-self-dual part of the (linearized) Riemann tensor and $\bar{t}$ is the scalar that lies in the same on-shell multiplet as $h^{++}$. Again amplitudes involving insertion of $(R^{+})^2\bar{t}$ and   $(R^{-})^2t$ will also encounter the same sub-leading  soft corrections as mentioned before. This would imply, among other things, that the two-loop four-point MHV amplitude will require corrections to $S^{(2)}_{\rm G}$ due to the presence of this term in the effective action, on top of those necessary due to the presence of IR-divergences.      
\section{Soft theorems for tree-level amplitudes in string theory}\label{string1}
In this section, we will discuss the soft theorem for superstring amplitudes. We will begin with explicit four and five-point examples in both open and closed string theories. After establishing the soft theorem for lower-point amplitudes, we will give a general argument based on BCFW recursion relations of string amplitudes. Furthermore in section~\ref{stringOPE}, we will present yet another independent analysis for the soft theorems in string amplitudes by the OPE of world-sheet vertex operators. 
\subsection{Soft theorem for open-string amplitudes: four and five-point example} \label{open45pt}
A general $n$-point color-ordered open string gluon amplitude at tree level can be expressed in terms of a basis of $(n{-}3)!$ functions~\cite{Mafra:2011nw, Mafra:2011nv},
\bea \label{generalopen}
\mathcal{A}(1,2, \ldots, n) = 
\sum_{\sigma \in S_{n-3}} F^{(2_{\sigma }, \ldots, (n{-}2)_{\sigma } )} 
A_{\rm YM} (1, 2_{\sigma }, \ldots, (n{-}2)_{\sigma }, n{-}1, n)
\eea
where multiple hypergeometric functions are given by
\bea
F^{( 2, \ldots, n{-}2)} = (-1)^{n-3} \int_{z_i < z_{i+1}} \prod^{n-2}_{j=2} dz_j \left( \prod |z_{i l}|^{s_{il}} \right)
\left(  \prod^{[n/2]}_{k=2} \sum^{k-1}_{m=1} {s_{mk}  \over z_{mk}} \right)
 \left( \prod^{n-2}_{k=[n/2]+1} \sum^{n-1}_{m=k+1} {s_{km}  \over z_{km}}   \right) \, ,\nonumber
\eea
where the Mandelstam variables are defined as $s_{ij}\equiv \alpha' (k_i{+}k_j)^2$. Here we have fixed SL$(2)$ symmetry by choosing $z_1=0, z_{n-1}=1$ and $z_n = \infty$. From the general expression (\ref{generalopen}), we find the four-point amplitude 
\bea \label{4pt1}
\mathcal{A}(1,2, 3, 4) = 
F^{(2)}A_{\rm YM}(1,2,3,4) \,, 
\eea
with 
\bea \label{4pt}
F^{(2)} = s_{12} \int^1_{0} dz_2 \, z^{s_{12}-1}_2 (1-z_2)^{s_{23}} = { \Gamma( 1 + s_{12} ) \Gamma( 1+ s_{23} )  \over \Gamma(1 + s_{12} + s_{23} )  } \, .
\eea
Use the fact that, in soft limit $k_2 \rightarrow \delta k_2 $ with $\delta \rightarrow 0$, 
\bea
{ \Gamma( 1 +  s_{12} ) \Gamma( 1+  s_{23} )  \over \Gamma(1 +  s_{12} +  s_{23} )  }
= 1 + \mathcal{O}( \delta^2 ) \, ,
\eea
it is easy to see that $\mathcal{A}(1, 2, 3, 4)$ satisfies the soft theorem, since $S_{\rm YM}^{(1)}(123)\mathcal{A}(134) =0$.

Let us now move on to the study of the soft limit for the five-point amplitude that can be written as
\bea  \label{5pt}
\mathcal{A}(1,2,3,4,5) 
=
F^{(2,3)} A_{\rm YM}(1,2,3,4,5) + F^{(3,2)} A_{\rm YM}(1,3,2,4,5) \, ,
\eea
where
\bea \label{integralrep}
F^{(2,3)} &=& s_{12} s_{34} \int^1_{0} dz_2 \int^1_{z_2} dz_3
z^{s_{12}-1}_2 z^{s_{13}}_3 z^{s_{23}}_{32} 
(1-z_2)^{s_{24}} (1-z_3)^{s_{34} -1 }   \, , \cr
F^{(3,2)} &=& s_{13} s_{24} \int^1_{0} dz_2 \int^1_{z_2} dz_3
z^{s_{12}}_2 z^{s_{13}-1}_3 z^{s_{23}}_{32} 
(1-z_2)^{s_{24}-1} (1-z_3)^{s_{34} } \, ,
\eea
with $z_{32} = z_3 - z_2$.

In $D=4$, we can take $k_{n{-}2}= k_3$ to be soft and solve for $\tilde{\lambda}_4$ and $\tilde{\lambda}_5$ using momentum conservation, 
\bea \label{k451}
\tilde{\lambda}_4
=
{\langle 5| (1+2) \over \langle 4 5 \rangle }  +\delta { \langle 5| 3 \over \langle 4 5 \rangle }  \, ,
\quad
\tilde{\lambda}_5
= 
{\langle 4| (1+2) \over \langle 5 4 \rangle }  +\delta { \langle 4| 3  \over \langle 5 4 \rangle }  \, ,
\eea
from which we can conveniently define
\bea \label{k452}
k'_4 &=& { |4 \rangle \langle 5| (1+2) \over \langle 4 5 \rangle }  \, ,  \quad
p_4 = { |4 \rangle \langle 5| 3 \over \langle 4 5 \rangle }  \, , \\
k'_5 &=& { |5 \rangle \langle 4| (1+2) \over \langle 54 \rangle }  \, ,  \quad
p_5 = { |5 \rangle \langle 4| 3 \over \langle 54 \rangle } \, .
\eea
Integrating over $z_3$ and keeping terms up to sub-leading  order we obtain
\bea\label{logbehav}
F_{S}^{(2,3)} =
s_{12} \int^1_{0} dz_2 \,
z^{s_{12}-1}_2 (1-z_2)^{s_{24'}}
\left[ 1  + \delta ( s_{23} + s_{34'} + s_{2p_4}  )  \log(1-z_2)     \right]
 \, . 
\eea
The leading term simply gives $F(1,2,4', 5')$, that appears in the four-point amplitude, and leads to \footnote{The leading soft-limit term for $n$-point amplitudes was analysed in~\cite{Mafra:2011nw}.}
\bea
{1 \over \delta^2} S^{(0)}_{\rm YM}(234) \mathcal{A}(1,2,4',5') \, .
\eea
Whereas the sub-leading  term, denoted by $F_{S^{(1)}}^{(2,3)}$, reads
\bea \label{F1sub}
F_{S^{(1)}}^{(2,3)} =
 { \langle 34 \rangle \langle 51\rangle [31] \over \langle 45 \rangle } s_{12}   
\int^1_{0} dz_2 \, z^{s_{12}-1}_2 (1-z_2)^{s_{24'}} \log(1-z_2) \, ,
\eea
here the identity $s_{23} + s_{34'} + s_{2p_4} =  { \langle 34 \rangle \langle 51\rangle [31] \over \langle 45 \rangle }$ has been used. The above integral can be computed straight-forwardly, however this is not necessary for our purposes as we will compare its expression with $S^{(1)}_{\rm YM}(234) \mathcal{A}(1,2,4',5')$ at the level of integrands. Similar consideration applies to $F^{(3,2)}$, which has a sub-leading contribution only, given by
\bea \label{F2sub}
F_{S^{(1)}}^{(3,2)} =
- s_{13} s_{24'} \int^1_{0} dz_2 z_2^{s_{12}} (1-z_2)^{s_{24'} - 1} \log(z_2) \, .
\eea
Combining the two contributions and expanding $\mathcal{A}(1,2,3,4',5')$, we find the sub-leading term  
\bea
A_{\rm YM}(1,2,4',5'){1 \over \delta} \left( 
 {\langle 24\rangle \over \langle 23 \rangle \langle 34\rangle }  F_{S^{(1)}}^{(2,3)} +  
 {\langle 12\rangle \over \langle 13 \rangle \langle 32\rangle }  F_{S^{(1)}}^{(3,2)} \right)
\eea
Now we are ready to compare this with the result of the soft operator acting on the four-point string amplitude, 
\bea
S^{(1)}_{\rm YM}(234) \mathcal{A}(1,2,4',5') 
&=&
\left( 
{1 \over \langle 23 \rangle } \tilde{\lambda}_3 \cdot  { \partial \over \partial \tilde{\lambda}_2 } 
+ {1 \over \langle 3 4 \rangle } \tilde{\lambda}_3 \cdot { \partial \over \partial \tilde{\lambda}_4 }
\right)
\mathcal{A}(1,2,4',5') \cr
&=&
\left( 
{\langle 24 \rangle \langle 51 \rangle [31] \over \langle 23 \rangle \langle 45 \rangle }   { \partial \over \partial s_{24'} } 
+ {\langle 12 \rangle [13] \over \langle 23 \rangle }  { \partial \over \partial s_{12} } 
\right)
\mathcal{A}(1,2,4',5') \, ,
\eea
where it is understood that $\tilde{\lambda}_4$ and $\tilde{\lambda}_5$ are solved by momentum conservation, and thus the result of the action of ${ \partial \over \partial \tilde{\lambda}_4 }$ on the amplitude vanishes. Now it is straightforward to see that
\bea
{\langle 24\rangle \over \langle 23 \rangle \langle 34\rangle }  F_{S^{(1)}}^{(2,3)} 
&=&{\langle 24 \rangle \langle 51 \rangle [31] \over \langle 23 \rangle \langle 45 \rangle }   { \partial \over \partial s_{24'} } F^{(2)}(1,2,4',5') \cr
{\langle 12\rangle \over \langle 13 \rangle \langle 32\rangle }  F_{S^{(1)}}^{(3,2)} 
&=&{\langle 12 \rangle [13] \over \langle 23 \rangle }  { \partial \over \partial s_{12} }  F^{(2)}(1,2,4',5') \, ,
\eea
where $F^{(2)}(1,2,4',5')$ is given in (\ref{4pt}). In order to check the validity of the second line in the above equation, it is convenient to use
\bea
F^{(2)}(1,2,4',5') = s_{24'} \int^1_{0} dz_2 z^{s_{12}}_2 (1 - z_2)^{s_{24'}-1} \, . 
\eea
This thus establishes the soft theorem for the five-point open superstring amplitude. Similar direct analysis can be applied to higher-point amplitudes, we have checked analytically that (\ref{generalopen}) satisfies the soft theorem for six points, see Appendix \ref{Appendix:sixpt}. 

\subsection{Soft theorem for closed-string amplitudes: four and five-point examples}\label{45ptclosed}
The tree-level closed-string amplitude can be written in terms of open-string tree amplitudes via KLT relations~\cite{KLTref, Paris},
\ba\label{KLT}
\mathcal{M}_n&=&
\pi^{3-n}\mathcal{A}_n(1,2,\ldots,n) \sum_{\{i\},\{j\}} f(i_1,\ldots,i_{\lfloor \frac n 2\rfloor{-}1}) \bar f(j_1,\ldots,j_{\lfloor \frac n 2\rfloor{-}2}) ]\mathcal{A}_n (\{i\}, 1,n{-}1, \{j\}, n)\nl
&+&  {\rm Perm}(2,\ldots,n{-}2)
\ea
where the sum inside the bracket is over $\{i\}\in {\rm Perm}(2,\ldots,\lfloor \frac n 2\rfloor)$, $\{j\}\in {\rm Perm} (\lfloor \frac n 2\rfloor{+}1,\ldots,n{-}2)$, and the functions $f$ and $\bar f$ are defined as
\ba
f(i_1,\ldots,i_m)&=&\sin(\pi s_{1\,i_m}) \prod_{k=1}^{m{-}1} \sin \pi \left( s_{1\,i_k}{+} \sum_{l=k{+}1}^m g(i_k,i_l)\right),\nl
\bar f (j_1,\ldots,j_m)&=& \sin(\pi s_{j_1\,n{-}1}) \prod_{k=2}^m \sin \pi \left( s_{j_k\,n{-}1} {+}\sum_{l=1}^{k{-}1} g(j_l, j_k)\right),
\ea
with $g(i, j)=s_{i j}$ for $i>j$ and $0$ otherwise. For four points, we have
\bea
\mathcal{M}_4(\{1,2,3,4\}) =\pi^{-1} \sin ( \pi s_{12} ) \mathcal{A}_4(1,2,3,4)\mathcal{A}_4(2,1,3,4) \, .
\eea
Consider the soft limit $k_2 \rightarrow \delta k_2, \, \delta \rightarrow 0$, we find 
\bea
\mathcal{M}_4(\{1,2,3,4\})\big{|}_{\rm div} &=&
{1 \over \delta^3 }S^{(0)}_{\rm YM}(123) S^{(0)}_{\rm YM}(421)  \mathcal{A}^2_3(1,3,4) 
\left( s_{12}  - \delta^2 \zeta_2 s_{12}^2 (s_{12} + s_{23} + s_{24})  \right) \cr
&=& 
{1 \over \delta^3 }s_{12} S^{(0)}_{\rm YM}(123) S^{(0)}_{\rm YM}(421)  \mathcal{M}_3(1,3,4) 
 \, ,
\eea
where momentum conservation has be used in the last step. Thanks to 
\bea
s_{12} S^{(0)}_{\rm YM}(123) S^{(0)}_{\rm YM}(421) =S^{(0)}_{\rm G}(2),\quad S^{(1)}_{\rm G}(2) \mathcal{M}_3(1,3,4) = S^{(2)}_{\rm G}(2)  \mathcal{M}_3(1,3,4) =0 \, ,\nonumber
\eea
we find that the closed string four-point amplitude satisfies the soft theorem. 

We then study the closed string amplitude at five points, which again can be expressed via KLT relations
\ba\label{KLT5}
\mathcal{M}_5(\{1,2,3,4,5\})&=& \pi^{-2} \big( \mathcal{A}_5(1,2,3,4,5) \mathcal{A}_5(2,1,4,3,5)\sin(\pi s_{12}) \sin(\pi s_{34})\nl
 &+& \mathcal{A}_5(1,3,2,4,5) \mathcal{A}_5(3,1,4,2,5)\sin(\pi s_{13}) \sin(\pi s_{24}) \big).
\ea
We will take leg $3$ to be soft, and with four-dimensional kinematics solve $\tilde{\lambda}_4$ and $\tilde{\lambda}_5$ using momentum conservation, with $k'_4, k'_5$ and $p_4, p_5$ defined as in (\ref{k452}). 

At the leading order, we have $\sin(\pi s_{3 i})=  \pi s_{3 i}{+}\mathcal{O}(\delta^3)$, and using the leading soft-theorem for open-string amplitudes we have (if we take the holomorphic limit)
\ba\label{leading5}
\mathcal{M}_5&=&\delta^{-3} s_{34'} S^{(0)}_{\rm YM}(2,3,4') S^{(0)}_{\rm YM}(4',3,5') \left[\pi^{-1} \sin(\pi s_{12}) \mathcal{A}_4(1,2,4',5')\mathcal{A}_4(2,1,4',5')\right]\nl
&+&\delta^{-3} s_{13} S^{(0)}_{\rm YM}(1,3,2) S^{(0)}_{\rm YM}(5',3,1)\left[\pi^{-1} \sin(\pi s_{24'}) \mathcal{A}_4(1,2,4',5')\mathcal{A}_4(4',2,5',1)\right]+\mathcal{O}(\delta^{-2}),\nl
\ea
where we recognize that the two combinations inside square brackets are two KLT representations of the same four-point amplitude, $\mathcal{M}_4(\{1,2,4',5'\})$, and the prefactors combine to the leading gravity soft-factor
\be
S^{(0)}_{\rm G}(3)=\sum_{i=1}^5\frac{[3i]}{\l 3i\r}\frac{\l xi\r\l yi\r}{\l x3\r\l y3\r}=\sum_{i=1,4} s_{3 i} \frac{\l i 2\r\l i 5\r}{\l i 3\r^2\l 3 2\r\l 3 5\r },
\ee
where we have used the four-dimensional form of $S^{(0)}_G$ and the gauge choice choose $x=2,y=5$. 

The sub-leading  order of eq.~(\ref{KLT5}) receives contribution from the sub-leading  order of $\mathcal{A}_5$'s: for the first term, we have $\frac{\partial}{\partial\tilde\lambda_2}$ in $S^{(1)}_{\rm YM} (2,3,4')  \mathcal{A}_4 (1,2,4',5')$, and for the second term, $\frac{\partial}{\partial\tilde\lambda_{1,2}}$ in $S^{(1)}_{\rm YM} (1,3,2)  \mathcal{A}_4 (1,2,4',5')$ and $\frac{\partial}{\partial\tilde\lambda_1}$ in $S^{(1)}_{\rm YM} (5',3,1) \mathcal{A}_4(4',2,5',1)$. Combining these terms and the sub-leading term from $\sin(\pi s_{2 4})=\sin(\pi s_{2 4'})+\delta \pi \cos(\pi s_{24'}) s_{2 p_4}$, we find
\ba \label{5ptclose}
\mathcal{M}_5|_{\mathcal{O}(\delta^{-2})}=&&\pi^{-1} \frac 1 {\l 2  3\r}\tilde\lambda_3 \cdot \frac{\partial \mathcal{A}_4(1,2,4',5')}{\partial\tilde\lambda_2}\left(\sin(\pi s_{1 2})\frac{[3 4']\l 4'  5'\r}{\l 3 5'\r} \mathcal{A}_4(2,1,4',5')\right.\nl
&&\left.-\sin(\pi s_{2 4'}) \frac{[1 3]\l 5'  1\r}{\l 5'  3\r} \mathcal{A}_4(4',2,5',1)\right)\nl
&+& \pi^{-1} \sin(\pi s_{2 4'}) \frac 1{\l 1  3\r}\tilde\lambda_3 \cdot \frac{\partial \mathcal{A}_4(1,2,4',5')}{\partial\tilde\lambda_1} \frac{[1 3]\l 5'  1\r}{\l 5'  3\r} \mathcal{A}_4(4',2,5',1)\nl
&+& \pi^{-1} \sin(\pi s_{24'}) \frac 1{\l 1  3\r}\tilde\lambda_3 \cdot \frac{\partial \mathcal{A}_4(4',2, 5',1)}{\partial\tilde\lambda_1} \frac{[1,3]\l 2  1\r}{\l 2  3\r} \mathcal{A}_4(1,2,4',5')\nl
&-& \cos (\pi s_{24'})\tilde\lambda_3\cdot \frac{\partial s_{2, 4'}}{\partial \tilde\lambda_1} \frac{[1 3]\l 1  2\r}{\l 1 3\r\l 3  2\r} \mathcal{A}_4(1,2,4',5') \mathcal{A}_4(4',2,5',1).
\ea
where on the last line we have rewritten $s_{2p_4} S^{(0)}_{\rm YM} (132)S^{(0)}_{\rm YM}(5'31)$ as a derivative operator acting on $s_{24'}$.

Now we compare this with $S_G^{(1)}(3) M_4$, which is given by
\be
\frac{1}{2} \sum_{i=1,i\neq 3}^5 \frac{[3i]}{\l 3i\r} \left( \frac{\l xi\r}{\l x3\r}+\frac{\l yi\r}{\l y3 \r}\right)\tilde\lambda_3^{\dot{\alpha}}\frac{\partial}{\partial \tilde\lambda^{\dot\alpha}_i}M_4\, .
\ee
The crucial step in dealing with the big bracket in (\ref{5ptclose}) is the use of the monodromy relation 
\be
\sin(\pi s_{12})\mathcal{A}_4(2,1,4',5')=\sin(\pi s_{24'}) \mathcal{A}_4(4',2,5',1),
\ee 
in order to simplify it to $\frac{[3 2]\l 2 5'\r}{\l 5' 3\r}\sin(\pi s_{24'}) \mathcal{A}_4 (4',2,5',1)$. This in turn can be combined with the third line to produce $(S^{(1)}_{\rm G}(3) \mathcal{A}_4(1,2,4',5')) \sin(\pi s_{1,2}) \mathcal{A}_5(4',2,5',1)$ with the gauge-choice $x=y=5$. Since $S^{(1)}_{\rm G}(3)$ is gauge-invariant, we can make a different choice $x=y=2$,  
and in this form the result is simplify $S^{(1)}_{\rm G}(3)$ acting on the second KLT representation of $\mathcal{M}_4$ in eq.~(\ref{leading5}):
\ba
\mathcal{M}_5|_{\mathcal{O}(\delta^{-2})}&=&  \frac{[1 3]\l 1 2\r}{\l 1 3\r\l 3 2\r} \tilde\lambda_3 \cdot \frac{\partial}{\partial\tilde\lambda_1} [\pi^{-1} \sin(\pi s_{24'}) \mathcal{A}_4(1,2,4',5') \mathcal{A}_4(4',2,5',1)] \nl
&=& S^{(1)}_{\rm G}(3) \mathcal{M}_4(\{1,2,4',5'\}) \, .
\ea

Finally we move to the order $\mathcal{O}(\delta^{-1})$, where one needs to consider: the product of sub-leading  contributions from $\mathcal{A}_5$'s, the sub-sub-leading contribution from the $\sin$ factors, and the sub-sub-leading contribution from either of the $\mathcal{A}_5$'s. We have worked out all contributions analytically (the details can be found in Appendix \ref{Appendix:5ptclose}), and checked against $S^{(2)}_{\rm G}(3) \mathcal{M}_4(\{1,2,4',5'\})$ numerically, we found perfect agreement. 

Two comments regarding closed-string soft theorems are in order.  First, we believe that the pattern we observed in the proof for $S^{(0)}_G$ and $S^{(1)}_G$ at five-points can be generalized to higher points. It would be desirable to explicitly check these first two orders of the soft graviton theorem, by KLT relations and repeated use of monodromy relations. 

Besides, we want to stress that the agreement at sub-sub-leading  order, unlike the first two orders, is not a direct consequence of KLT and monodromy relations. In particular, in KLT representation it involves non-universal sub-sub-leading  soft behavior of open-string amplitudes, and it would be interesting to understand better how they combine nicely into the universal $S^{(2)}_G$ acting on the lower-point amplitude. 
  
\subsection{Soft theorems of string amplitudes from BCFW recursion relations} \label{BCFWstring}
In this section we will give a general argument for the soft theorems in string theories based on BCFW recursion relations. BCFW recursion relations for scattering amplitudes in filed theories~\cite{Britto:2004ap, Britto:2005fq} have been generalized to open- and closed-string amplitudes~\cite{Cheung:2010vn, Boels:2010bv}\footnote{We are aware that the recursion relation has only been explicitly checked to be correct for a few examples.}. For instance for the color-ordered open string amplitudes, one has
\bea \label{stringBCFW}
\mathcal{A}(1, 2, \ldots , n{-}1, n ) 
= \sum_i \sum_{\rm states \, I} 
\mathcal{A}_{L}( \hat{1}, 2, \ldots, i, I ) {1 \over k^2_I + m^2_I } 
\mathcal{A}_R( -I, i{+}1,  \ldots, \hat{n} ) \, .
\eea
In practice, since the sum runs over an infinite number of states, the recursion may not be so useful for computing scattering amplitudes in string theories (See papers~\cite{Chang:2012qs, Boels:2014dka} for recent development on application of BCFW recursion relations in string amplitudes.). However, the above recursion relation is very useful for our purpose of proving the soft theorems. Here we take holomorphic soft limit on leg $1$. First of all, for the terms with $i>2$ in the recursion relation (\ref{stringBCFW}) are regular, just as the recursion relations for field theories. As for the case when $i=2$, the crucial observation is that only massless states can contribute to the soft limit, since the singularity arises from ${1 \over k^2_I + m^2_I }$. Thanks to the recursion relation, in the soft limit, the divergent part of an open superstring amplitude reduces to 
\bea
\mathcal{A}(1, 2, \ldots , n{-}1, n ){\Big{|}_{\rm div}} 
=  
\mathcal{A}_{3}( \hat{1}, 2,  I ) {1 \over k^2_I  } 
\mathcal{A}_{n-1}( -I, 3,  \ldots, \hat{n} ) \, ,
\eea
note that the internal state is a massless gluon now. Since the three-point open superstring amplitude is identical to the one in  SYM, we see that the result of this particular BCFW channel takes the same form as for Yang-Mills amplitudes, {\it i.e.} eq.(\ref{BCFWYM}), 
\bea
\mathcal{A}(1, 2, \ldots , n{-}1, n ){\Big{|}_{\rm div}} 
=  
\left( {1 \over \delta^2 } S^{(0)}_{\rm YM}(n12) + {1 \over \delta } S^{(1)}_{\rm YM}(n12) \right)
\mathcal{A}_{n-1}( 2 , 3,  \ldots, n) \, ,
\eea
which are universal parts of the amplitude. The same argument applies to closed superstring amplitudes.  

The BCFW argument can also apply to bosonic string amplitudes. For the case of open strings, the conclusion is the same since there is no other massless state, except for the gluon. Whereas for bosonic closed string amplitudes, besides the graviton we have also the massless dilaton (Kalb-Ramond field does not contribute since there is no three-point amplitude with two gravitons and a Kalb-Ramond field), which could contribute to $\mathcal{M}_{n}{\big{|}_{\rm div}}$. The contribution of the dilaton $\phi$ is of order $\mathcal{O}(\delta^{-1})$, and spoils the $S^{(2)}_{\rm G}\mathcal{M}_{n-1}$ term by a factor of
\bea
\mathcal{M}^{\phi}(1^+, 2, \ldots , n{-}1, n ){\big{|}_{\rm div}} 
&=&  
\sum_{i} \mathcal{M}_{3}( \hat{1}^+, i^+,  I ) {1 \over k^2_I  } 
\mathcal{M}_{n-1}( -I, 3,  \ldots, \hat{n} ) \cr
&=&
-{2 \over \delta} \sum_i {[1i]^3 \over \langle 1i \rangle } 
\mathcal{M}_{n-1}( \phi, 3,  \ldots, \hat{n} ) \, ,
\eea
where we have emphasized the fact that only the amplitude with helicity $( h^{++}, h^{++}, \phi )$ (or its conjugate) is non-vanishing by making helicity dependence explicit.

\section{Soft limit of superstring amplitudes: world-sheet analysis} \label{stringOPE}

We here discuss how to derive soft theorems for string amplitudes from the perspective of world-sheet OPE in the NS-R approach. The analysis can be systematised and even in principle one can derive further sub-leading terms and investigate their universality. 
\subsection{Preliminaries}
The Euclidean world-sheet is parameterized by the coordinates $z=e^w$, $w=\tau+i\sigma$, where for open strings $\sigma\in [0,\pi]$, $\tau \in (-\infty, +\infty)$, while for closed strings we have $\sigma\in [0,2\pi]$, $\tau \in (-\infty, +\infty)$. For convenience, we will use units such as $2\alpha^\prime = 1$ for open strings and $\alpha^\prime = 2$ for closed strings \cite{Kiritsis:2007zza}.

We will analyze both the bosonic string and the superstring. For the open bosonic string, the vertex operator for a massless vector boson
is
\eq
V_A=(\epsilon{\cdot} \partial X)e^{ikX} 
\eqe
where $k^2=\epsilon\cdot k =0$. 
Similarly, for the closed bosonic string, the graviton vertex operator is 
\eq
V_G=E_{\mu\nu} \partial X^\mu \bar{\partial} X^\nu e^{ikX}
\eqe
where  $E_{\mu\nu}= E_{\nu\mu}$, $k^2=k^\mu E_{\mu\nu} = g^{\mu\nu} E_{\nu\mu} = 0$. In explicit computations, it is often convenient to set $E_{\mu\nu} = \epsilon_\mu \epsilon_\nu$ and factorise the vertex into two chiral parts. 

In the Neveu-Schwarz (NS) sector of the superstring, the vertex operator for a gauge boson in the (-1) super-ghost picture is
\eq
V^{({-}1)}_A =( \epsilon{\cdot}\psi) e^{-\varphi} e^{ikX},\quad  \eqe
where $\varphi$ is the boson for the super-ghosts. For the graviton one has 
\eq
V_G^{({-}1,{-}1)} = E_{\mu\nu}\psi^\mu \tilde\psi^\nu  e^{-\varphi}  e^{-\tilde\varphi} e^{ikX}
\eqe
The vertex operators in the (0) picture are:
\bea
V_A^{(0)} = (i\epsilon {\cdot} \partial X + k{\cdot} \psi\, \epsilon{\cdot}\psi) e^{ikX}
\eea
and
\bea
V_G^{(0,0)} =E_{\mu\nu}(i\partial X^\mu + k{\cdot} \psi\, \psi^\mu)(i\bar\partial X^\mu + k{\cdot} \tilde\psi\, \tilde\psi^\mu) e^{ikX}\,.
\eea
We will use the following normalization for the correlators:
\eq
\langle X^\mu(z_1)X^\nu(z_2)\rangle=- {\alpha'}g^{\mu\nu}\ln |z_1{-}z_2|^2,\quad\langle \psi^\mu(z_1)\psi^\nu(z_2)\rangle=\frac{g^{\mu\nu}}{z_1{-}z_2}\,.
\eqe

In the following, we will need the generators of the Lorentz group. In the open bosonic strings they are 
\be
J^{\mu\nu} = {1\over \pi }\int_0^\pi d\sigma [ X^\mu\partial_\tau X^\nu  -  X^\nu\partial_\tau X^\mu ] \,,
\ee
while for the open superstring  in the $q=0$ super ghost picture, we have:
\be
J^{\mu\nu}_{(0)}= {1\over \pi }\int_0^\pi d\sigma [ X^\mu\partial_\tau X^\nu  -  X^\nu\partial_\tau X^\mu + \psi^\mu \psi^\nu ] \,.
\ee
The commutator of $J^{\mu\nu}$ with the gauge boson vertex operator takes the form:
\be
[J_{\mu\nu}, V_A(k)] = \left(\epsilon_{[\mu} {\partial \over\partial \epsilon^{\nu]}}+ k_{[\mu} {\partial \over\partial k^{\nu]}} \right) V_A(k)
\ee
This analysis extends directly to the open superstring (or the other open fermionic strings) and to the closed bosonic and super- (or fermionic) strings. In the latter cases one should keep in mind that there is a single conserved center of mass momentum $P^\mu = p^\mu_0$ and a single conserved angular momentum
\be
J_{cl}^{\mu\nu} =  x_0^\mu p_0^\nu -  x_0^\nu p_0^\mu + \hat{J}_L^{\mu\nu} + \hat{J}_R^{\mu\nu}
\ee
where $\hat{J}_{L,R}^{\mu\nu} $ denote the contribution of the oscillators including fermionic zero-modes $\psi^\mu_0 \psi^\nu_0$ or  $\bar\psi^\mu_0 \bar\psi^\nu_0$ when present (Ramond sector of the superstring). With some effort one can check that 
\be
[J_{cl\,\mu\nu}, V_{G}(k)] = \left(2\epsilon_{[\mu} {\partial \over\partial \epsilon^{\nu]}}+ {k_{[\mu}}{\partial \over\partial k^{\nu]}} \right) V_G(k)\,.
\ee
for the graviton with $E_{\mu\nu} = \epsilon_\mu \epsilon_\nu$.
An important property that will be relevant to our discussion is that $J^{\mu\nu}$ is BRST invariant, and thus the commutator of $V$ and $J$ remains BRST invariant.   
Note also that the leading term in the gluon vertex operator contains the world-sheet current ${\cal J}_P^\mu = \partial_z X^\mu = \partial_{\tau} X^\mu = \Pi^\mu$ (momentum conjugate to $X^\mu$) for the space-time momentum operator $P^\mu$, while the sub-leading term contains the  world-sheet current ${\cal J}_J^{\mu\nu} = X^\mu\partial_z X^\nu - X^\mu\partial_z X^\mu + \psi^\mu \psi^\nu$ for angular momentum $J^{\mu\nu}$.

This is in line with the fact that the on-shell vertex operator for a massless vector at $k=0$ {\it i.e.} with a constant field-strength is precisely $V_F= F^{\mu\nu}  \int dz [X_\mu \partial X_\nu - X_\nu \partial X_\mu + \psi_\mu\psi_\nu]$. Indeed, when $V_F$ is inserted in the action it changes the boundary conditions from $X_\mu\partial_\sigma X^\mu \vert_{ \sigma = 0,\pi}= 0$ to $X_\mu\partial_\sigma X^\mu\vert_{ \sigma = 0,\pi}= X_\mu F^\mu{}_\nu \partial_\tau X^\nu \vert_{ \sigma = 0,\pi}$ and similarly for fermions (when present).

\subsection{Open superstring amplitudes on the disk }
Color ordered disk amplitudes are given by: 
\bea
\cA(1,2,\ldots ,n) = i g_s^{n-2} \int_{0\le z_2\le \ldots z_{n-2}\le 1} dz_2 \ldots dz_{n-2}  \langle cV(1) V(2) \ldots cV(n-1) cV(n)\rangle
\eea
where $V$ denote the vertex operators and $c$ the conformal ghost. In order to saturate the super-ghost charge one needs $\sum_i q_i = -2$. This can be satisfied taking two vertices in the $q=-1$ picture and the remaining $n-2$ in the $q=0$ picture. In order to make the analysis of the soft limit transparent, it is convenient to take the vertex that goes `soft' in the $q=0$ picture and the two neighboring ones in the $q=-1$ picture. We will follow our previous convention where the soft leg is in the last position labelled by $n{+}1$.  

We now consider the OPE between the soft vertex $V_A^{(0)}$ and its adjacent vertices $V_A^{(-1)}$ at $z_{1}$ and $z_{n}$:
\bea\label{OpenOPE}
V_A^{(0)}(z_s) V_A^{(-1)} (z_{n}) \approx |z_s{-}z_{n}|^{k_s {\cdot} k_{n} - 1} e^{-\varphi(z_{n})} e^{i (k_s+k_{n})X(z_{n}) } \nonumber\\
\times\left({\epsilon_s {\cdot} k_{n}\,  \epsilon_{n}{\cdot}\psi} - \epsilon_{n}{\cdot} k_s \, \epsilon_s{\cdot}\psi + \epsilon_{n}{\cdot} \epsilon_s\, k_s{\cdot} \psi\right) (z_{n})+  \ldots
\eea
where $\ldots$ indicate terms sub-leading  in $|z_s{-}z_{n}|$. The integral over $z_s$ can be done using the identity\footnote{This is a consequence of $\delta(x)=\lim_{s\rightarrow 0}sx^{s-1}$} 
\eq
\int_0^\epsilon x^{s-1}f(x)=\frac{f(0)}{s}+\mathcal{O}(s^0)\,,
\eqe
thus the leading term in the expansion of $k_s$ is simply $(\epsilon_s{\cdot}k_{n} /k_s{\cdot}k_{n})V_A^{(-1)}(n)$. 

At the next order, from the terms appearing in eq.(\ref{OpenOPE}) we obtain:
\eq
\frac{2}{k_s\cdot k_{n}}e^{-\varphi(z_{n})} e^{i k_{n}X(z_{n}) }
\left(i{\epsilon_s {\cdot} k_{n} \, \epsilon_{n}{\cdot}\psi\,  k_s{\cdot} X}+ \epsilon_{n}{\cdot} k_s\,  \epsilon_s{\cdot}\psi - \epsilon_{n}{\cdot} \epsilon_s \,k_s{\cdot} \psi\right) (z_{n})\,.
\eqe
The term proportional to $k_s{\cdot} X$ is responsible for the logarithms that appear in the explicit expansion of the amplitudes in the soft limit (see {\it e.g.} (\ref{logbehav}) ) and can be decomposed into a symmetric and anti-symmetric piece under the exchange $k_s\leftrightarrow \epsilon_s$. The symmetric piece is BRST exact. To see this note that the term we are interested in, $\epsilon_s {\cdot} k_{n}k_s{\cdot} X+k_s {\cdot} k_{n}\epsilon_s{\cdot} X$, can be written as:
\eq
\epsilon_{s\mu}k_{s\nu}X^{(\mu}k_{n}^{\nu)}=\frac{\epsilon_{s\mu}k_{s\nu}}{\pi}\int_0^\pi d\sigma \partial_{\tau} X^{(\mu}X^{\nu)}=\frac{\epsilon_{s\mu}k_{s\nu}}{\pi}\int_0^\pi d\sigma \{Q_{BRST}, bX^{\mu}X^{\nu}\}\,.
\eqe
where $b$ is the anti-ghost. Thus only the anti-symmetric piece is in the BRST cohomology. Putting everything together, we find that the sub-leading soft term is given by:
\eqa
\nonumber &&\frac{(F_s)_{\mu\nu}}{k_s\cdot k_{n}}
\left(i  {k_{n}^{\mu} X^{\nu} \epsilon_{n}{\cdot}\psi}+ \epsilon_{n}^\mu\psi^\nu \right)e^{-\varphi} e^{i (k_{n})X } (z_{n})\,\\
\nonumber &=&\frac{(F_s)_{\mu\nu}}{k_s\cdot k_{n}}
\left({k_{n}^{\mu}}\frac{\partial}{\partial k_{n\nu}}+ \epsilon_{n}^\mu\frac{\partial}{\partial \epsilon_{n\nu}} \right)V_A^{(-1)} (z_{n})\,,
\eqae
where $F_s\equiv k_{s[\mu}\epsilon_{s\nu]}$. In other words, the two terms combined neatly produce:
\be
 \frac{(F_s)_{\mu\nu}}{k_s\cdot k_{n}} [J^{\mu\nu},V_A^{(-1)} (z_n) ] 
 \ee
 where $J_{\mu\nu}$ is the total angular momentum, defined before, that acts on both polarisation (spin) and momentum (orbital). Thus we find that in the soft-limit, the sub-leading  contribution is given by the commutator of a BRST invariant operator with its adjacent vertex operators:
 \eq
 \frac{(F_s)_{\mu\nu}}{k_s\cdot k_{n}} \langle [J^{\mu\nu},V_A^{(-1)} (z_n) ]V_A^{(-1)}(z_1)\cdots \rangle - \frac{(F_s)_{\mu\nu}}{k_s\cdot k_{1}}\langle V_A^{(-1)}(z_n)[J^{\mu\nu},V_A^{(-1)}(z_1) ] \cdots\rangle\,.
 \eqe
 
Let us stress that the final results, derived with a specific choice of super-ghost pictures and position of the soft gluon, are very general and do not depend on these choices at all. 
In particular, had we chosen one of the `hard' vertices to be in the $q=0$ picture or the `soft' vertex to be in the $q=-1$ picture, the leading singularity in the OPE would have contained terms like
\be
|z_s-z|^{k_s{\cdot}k -2} \epsilon_s{\cdot}\epsilon \, e^{i(k_s+k)X} \times \left( {1\,\, {\rm or} \,\, e^{-2\varphi}} \right)
\ee
that would have not contributed to the leading term in the soft limit since it would have produced a `pole' $1/(k_s{\cdot}k -1)$ upon integration over $z_s$ around $z$. The sub-leading terms in the OPE such as
\be
|z_s-z|^{k_s{\cdot}k -1} e^{i(k_s+k)X} [\epsilon_s{\cdot}\epsilon (k_s-k){\cdot}\partial X + \epsilon_s{\cdot}\psi\,\epsilon{\cdot}\psi]\times \left( {1\,\, {\rm or}\,\, e^{-2\varphi}} \right)
\ee
would have then produced the desired `pole' $1/k_s{\cdot}k$ in the soft limit. With some effort, one could check that the leading and  sub-leading terms in the `soft' expansion be the same as in our analysis.

Moreover our analysis applies to superstring gluon amplitudes at tree level in any dimension $D\le 10$. Indeed, even after compactification the vertex operator for a massless gluon remains unchanged. One should simply restrict momentum and polarisation to have non-zero components only along the non-compact directions. In other words the vertex operator involves the `identity' operator of the CFT$_2$ governing the dynamics of the internal space. In particular, in $D=4$ there are only two physical polarisations and one can conveniently switch to  the spinor helicity basis, whereby a generic massless vector polarisation is the sum of plus and minus helicities. 
 
\subsection{Closed superstring amplitudes on the sphere}
In order to derive the behaviour of graviton (in fact any NS-NS massless state) amplitudes for closed superstrings on the sphere we start from the standard definition 
\bea
\cM(1,2,\ldots ,n) = i g_s^{2(n-2)} \int_{S^2} dz_2 \ldots dz_{n-2}  \langle c\bar{c}V(1) V(2) \ldots c\bar{c}V(n-1) c\bar{c}V(n)\rangle
\eea
where $V = V_L V_R$ denote closed-string vertex operators and $c$ the conformal ghost. 

As in the open superstring case, in order to saturate the super-ghost charge on the sphere one needs $\sum_i q_i = -2$  both for left- and right-movers. The simplest way to satisfy this condition is to take two vertices in the $q=-1$ picture and the remaining $n-2$ in the $q=0$ picture. In order to make the analysis of the soft limit transparent, it is convenient to take the closed-string vertex that becomes `soft' in the $q=0$ picture. 

In the soft limit, $k\rightarrow 0$, $V_G^{(0,0)}(z_s)$ becomes a total derivative and the integral over $z_s$ only receives contribution from the boundary points $z_s = z_i$, where the `soft' vertex  in the $q=0$ picture collides with non-soft ones. If the latter is in the $q=-1$, the result is completely determined by the OPE
\bea
V_G^{(0,0)}(z_s) V_{G}^{(-1,-1)} (z_i) \approx |z_s{-}z_i|^{2k_s{\cdot} k_i  - 2} e^{-\varphi(z_i) -\tilde\varphi(\bar{z}_i)} e^{i (k_s+k_i)X(z_i,\bar{z}_i )} \times
\nonumber\\
 (\tilde\epsilon_s {\cdot} k_{i}\,  \tilde\epsilon_{i}{\cdot}\psi - \tilde\epsilon_{i}{\cdot} \tilde{F}_s{\cdot}\tilde\psi )(\bar{z}_i) \left(\epsilon_s {\cdot} k_{i}\,  \epsilon_{i}{\cdot}\psi - \epsilon_{i}{\cdot} F_s{\cdot}\psi \right)(z_i) + \ldots\nonumber 
\eea
Integration over $z_s$ produces a pole $\pi/ k_s{\cdot} k_i $ from the most singular term in the OPE and, up to an overall  operator $e^{-\varphi(z_i) -\tilde\varphi(\bar{z}_i)} e^{i k_iX(z_i,\bar{z}_i )}$, the numerator can be expanded in $k_s$ as:
\eqa
\mathcal{O}(k_s^{0}):\; &&(\tilde\epsilon_s {\cdot} k_{i}\,  \tilde\epsilon_{i}{\cdot}\tilde\psi)(\epsilon_s {\cdot} k_{i}\,  \epsilon_{i}{\cdot}\psi)\nonumber\\
\mathcal{O}(k_s^1):\; &&\{i (k_s{\cdot} X)(\tilde\epsilon_s {\cdot} k_{i}\,  \tilde\epsilon_{i}{\cdot}\tilde\psi)(\epsilon_s {\cdot} k_{i}\,  \epsilon_{i}{\cdot}\psi)- \epsilon_s {\cdot} k_{i}\,  \epsilon_{i}{\cdot}\psi(\epsilon_{i}{\cdot} \tilde F_s{\cdot}\tilde\psi )\nonumber\\
&&- \tilde\epsilon_s {\cdot} k_{i}\,  \tilde\epsilon_{i}{\cdot}\tilde\psi(\epsilon_{i}{\cdot} F_s{\cdot}\psi )\}\nonumber\\
\mathcal{O}(k_s^2):\; &&\{i (k_s{\cdot} X)[(\tilde\epsilon_s {\cdot} k_{i}\,  \tilde\epsilon_{i}{\cdot}\tilde\psi) \epsilon_{i}{\cdot} F_s{\cdot}\psi+(\epsilon_s {\cdot} k_{i}\, \epsilon_{i}{\cdot}\psi) \tilde\epsilon_{i}{\cdot} \tilde{F}_s{\cdot}\tilde\psi]\nonumber\\
&&-(k_s{\cdot} X)^2(\tilde\epsilon_s {\cdot} k_{i}\,  \tilde\epsilon_{i}{\cdot}\tilde\psi)(\epsilon_s {\cdot} k_{i}\,  \epsilon_{i}{\cdot}\psi)/2+\tilde\epsilon_{i}{\cdot} \tilde{F}_s{\cdot}\tilde\psi \epsilon_{i}{\cdot} F_s{\cdot}\psi\} 
\eqae
At $\mathcal{O}(k_s^{0})$, this gives the leading soft behavior as:
\eq
\mathcal{O}(k_s^{-1}):\quad\pi\frac{(\tilde\epsilon_s {\cdot} k_{i})(\epsilon_s {\cdot} k_{i})}{k_s{\cdot} k_i }V_G^{({-}1,{-}1)}(z_i)\,.
\eqe
From the open string analysis, we have seen that it is convenient to rewrite the relevant terms in the form
\eq
\epsilon_{i}{\cdot} F_s{\cdot}\psi =F^{\mu\nu}_{s}\epsilon_{i\mu}\frac{\partial}{\partial \epsilon_i^{\nu}} \epsilon_i{\cdot}\psi\,,\quad i(X{\cdot} [k_s)(\tilde\epsilon_s] {\cdot} k_{i}\,  \tilde\epsilon_{i}{\cdot}\tilde\psi)e^{ik_i\cdot X}= \tilde{F}_{s}^{\mu\nu}k_{i\nu}\frac{\partial}{\partial k_i^{\mu}} \,\tilde\epsilon_{i}{\cdot}\tilde\psi\, e^{ik_i\cdot X}\,.
\eqe
Using these identifications and taking into account the symmetrization of the polarization vectors on leg $s$, for the sub-leading term we find,
\eqa
\mathcal{O}(k_s^{0}):\,\, && \frac{1}{2k_s{\cdot} k_i }
\Big[
(\epsilon_s {\cdot} k_{i}\,  \tilde\epsilon_{s}{\cdot}k_i)k_s^{\mu}\frac{\partial}{2\partial k_{i}^\mu}-(\epsilon_s {\cdot} k_{i}\, k_{s}{\cdot}k_i)\tilde{\epsilon}_s^{\mu}\frac{\partial}{2\partial k_{i}^\mu}- \epsilon_s {\cdot} k_{i}\, \tilde F_{s}^{\mu\nu}(\tilde\epsilon_{i\mu}{\cdot} \partial_{\tilde\epsilon_i^{\nu}} )\cr
&& + \,
(\epsilon\leftrightarrow \tilde{\epsilon})\Big]
V_G^{({-}1,{-}1)}(z_i)
=
\pi\frac{k_{i\mu}E_{s}^{\mu\rho}}{k_s{\cdot} k_i }
\left[
k_s^{\nu}J^{total}_{\rho\nu},V_G^{({-}1,{-}1)}(z_i)
\right]
\eqae 
where $J^{total}=J+\tilde{J}$ and $E_{s}^{\mu\nu}=\epsilon^{(\mu}\tilde{\epsilon}^{\nu)}/2$. Similar analysis for the sub-siub-leading order contribution yields:
\eqa
\mathcal{O}(k_s^{1})&:&\quad\pi\frac{E_{s}^{\mu\nu}}{2k_s{\cdot} k_i }[k_s{\cdot} J^{total}_{\mu}k_s{\cdot} J^{total}_{\nu},V_G^{({-}1,{-}1)}(z_i)]\,.
\eqae 
Thus we see that by soft expanding the result of the OPE between the soft and hard-vertex operators, we recover the field theory soft theorem, written in BRST invariant operator language.

Finally, notice that if one replaces the `soft' graviton with a `soft' dilaton or a `soft' Kalb-Ramond B-field the leading term vanishes. It is well known that the soft-dilaton limit of the $n{+}1$-pt amplitude gives the derivative of the amplitude wrt the string tension \cite{Seiberg:1986bm, Mayr:1993vu}, since the zero-momentum dilaton vertex operator is essentially the world-sheet action. 

In general the dilaton in $D=10$ and the other moduli fields in lower dimensions are governed by a non-linear $\sigma$-model and decouple at zero momentum like soft pions. An $n+1$-point amplitude with a soft modulus field is finite and given by the sum of $n$ contributions that represent the derivative withe respect to the constant VEV of the modulus field of the $n$-point amplitude without modulus field. Following this line of argument, many threshold corrections to (higher-derivative) terms in the effective superstring actions have been computed. See e.g. \cite{Kiritsis:2007zza} for a pedagogical presentation and references therein.

A slightly different story can be told for the insertion of a soft dilaton in the bulk of a disk with open string insertions on the boundary. The soft dilaton tadpole captures the divergence of the loop amplitude on a cylinder in the limit where it becomes infinitely long and thin. This divergence studied in detail in the early days of `dual' models \cite{Ademollo:1975pf} is absent in any consistent superstring background since it is related by super-symmetry to tadpoles in the R-R sector which, in turn, cancel in anomaly-free theories \cite{Bianchi:2000de}.

\section{Conclusions} 

In this paper we addressed two questions regarding soft gluon and graviton theorems. (1) Can we find representations of loop-integrands that manifestly satisfied tree-level soft theorems? (2) Are the tree-level soft theorems protected unmodified for effective theories with higher-dimensional operators or string theory at finite $\alpha'$? Concerning (1), we have found that for planar $\mathcal{N}=4$ SYM, the momentum twistor representation derived from loop-level BCFW recursion indeed manifests the soft behavior dictated by the unrenormalized (tree) soft-theorem. Similar conclusion can be arrived for one-loop amplitudes for $\mathcal{N}<4$ SYM in the CSW representation. For (2), we found that soft theorems are respected in a wide range of effective field theories, even for those with $F^3$ or $R^3$ interaction vertices; more importantly, they hold for open and closed superstring tree-level amplitudes, as verified by explicit computations, as well as general analysis based on BCFW recursion relations and world-sheet OPE. However, the sub-sub-leading soft graviton theorem is modified at tree level for theories with $R^2\phi$ vertex, and for bosonic closed-string theory. Note that while $R^2\phi$ interaction terms can be suppressed at tree level via supersymmetry, it can be generated by U(1) anomalies for $\mathcal{N}\leq4$ supergravity theories.

One interesting further direction would be turning (1) into a constructive way of constraining the form of loop integrands in more general theories. We have seen that only those exact integrands in planar SYM exhibit manifest soft behaviour identical to that of tree-level amplitudes; for other cases, including pure Yang-Mills and gravity theories, we do not know any form of the integrands that manifest soft theorems, but it is very likely that such integrands do exist. We expect soft theorems to be extremely useful in the search for these new representations of loop amplitudes. As discussed in~\cite{FreddyEllis}, it can be worthwhile to interpret not only soft limit but also collinear and factorization limits for loop amplitudes as kinematic limits to be be taken before expanding in regulators. In this way loop integrands behave very similar to tree-level amplitudes, as we can see from the BCFW-like recursion in $\mathcal{N}=4$. It would be fascinating to explore other formulations of loop integrands resembling those at tree level (e.g. twistor-string~\cite{Witten:2003nn} or scattering-equation~\cite{CHY} formulas), in $\mathcal{N}=4$ and beyond, based on their behavior in such kinematic limits. 

For integrated soft theorems, we have shown that loop corrections can be easily understood via the presence of symmetry anomalies, in particular conformal anomalies. Note that we have only used the conformal anomaly associated with generic kinematics, whose analytic form is not well known. On the other hand, the conformal anomaly associated with collinear kinematics is well studied, and thus it will be interesting to work out what constraints do these collinear anomalies impose. Finally, the fact that gluon soft theorems for all-plus amplitude is not renormalized, can be associated with conformal symmetry being unbroken at loop level for self-dual Yang-Mills. Similarly the all-plus amplitude for gravity is also unrenormalized. Might there be some hidden symmetry for tree-level gravity amplitudes that is respected at loop level for self-dual gravity, such that the soft theorems are protected?  

It is highly desirable to generalize our investigations to string amplitudes with higher-genus. In this respect, it is quite remarkable that the BCFW-like recursion relation (\ref{AHBCFWrecrel}) derived in \cite{ArkaniHamed:2010kv} closely resembles the three boundary contributions (pinching limits) of the world-sheet moduli space of an string amplitude at higher genus. The first corresponding to the collision of two external vertices. The second to the factorisation into two lower genus amplitudes (separating tube). The third to the degeneration of a tube/strip (pinching cycle). This analogy strongly suggests that, at least in the maximally supersymmetric case, superstring loop amplitudes should satisfy the same soft theorems as at tree level. It would also be interesting to further investigate the role of `soft dilaton' limits in the renormalization of the string tension and coupling constant. 

\section*{Acknowledgements}
It is a pleasure to thank Zvi Bern, Henrik Johansson for private communications, and Andreas Brandhuber, Radu Roiban, Rodolfo Russo, Gabriele Travaglini and Brian Wecht for helpful discussions.  The work of S. H is supported by Zurich Financial Services Membership and the Ambrose Monell Foundation. The work of Y-t. H is supported by the National Science Foundation Grant PHY-1314311. The work of C.W is supported by the Science and Technology Facilities Council Consolidated Grant ST/J000469/1 {\it String theory, gauge theory \& duality.}. The work of M.~B. is partially supported by the ERC Advanced Grant n. 226455 {\it Superfields} and was initiated while M.~B. was at QMUL holding a Leverhulme Visiting Professorship.

\appendix
\section{Symmetry constraints on soft functions}\label{sec:SuperSoft}
Here, we will derive the super-soft functions using the special SUSY generator $\mathfrak{S}_{Aa}=\sum_i\frac{\partial^2}{\partial \lambda_i^a\partial \eta_i^A}$, which holds classically for super Yang-Mills theory. Again we impose 
\eq
\left(\mathfrak{S}_{0}+\frac{1}{\delta}\mathfrak{S}_{s}\right)(\frac{1}{\delta^2}\mathcal{S}^{(0)}A_n+\frac{1}{\delta}\mathcal{S}^{(1)}A_n)=0
\eqe
We will begin with the well known result that $\mathcal{S}^{(0)}=S^{(0)}$, then order $\delta^{-3}$ is trivially satisfied. For $\delta^{-2}$ we have the following constraint:
\eq
\mathfrak{S}_0S^{(0)}A_n+\mathfrak{S}_s\mathcal{S}^{(1)}A_n=-\left(\frac{\lambda_n}{\la ns\ra^2}\frac{\partial}{\partial \eta_n}+\frac{\lambda_1}{\la 1s\ra^2}\frac{\partial}{\partial \eta_1}\right)A_n+\mathfrak{S}_s\mathcal{S}^{(1)}A_n=0\,.
\eqe
Now acting $\mathfrak{S}_s$ on the bosonic part of $\mathcal{S}^{(1)}$ gives 0, thus in order for the above equation to hold, one must include a fermionic term. Again going through the same analysis, one finds that the requisite fermionic piece is given by:
\eq
\frac{\eta_s}{\la s1\ra}\frac{\partial}{\partial\eta_1}+\frac{\eta_s}{\la sn\ra}\frac{\partial}{\partial\eta_n}\,.
\eqe
Thus we see that the supersymmetrized soft function is given by:
\eq\label{X3}
\mathcal{S}^{ (1)}=\mathcal{S}^{ (0)}\left[ { \langle s  n \rangle \over \langle 1 n\rangle } \left( \tilde{\lambda}_s \cdot { \partial \over \partial
 \tilde{\lambda}_1 } + \eta_s \cdot  { \partial \over \partial \eta_1  }   \right)+
 { \langle s  1 \rangle \over \langle n  1\rangle } 
 \left( \tilde{\lambda}_s \cdot { \partial \over \partial
 \tilde{\lambda}_n } + \eta_s \cdot  { \partial \over \partial \eta_n  }   \right) \right]
\eqe
that is exactly what was found in ~\cite{HHW} via recursion relations. 
\section{Soft theorem for six-point open string amplitude} \label{Appendix:sixpt}
The six-point open superstring amplitude can be expressed in terms of $(6-3)!=6$ YM amplitudes and as many multiple hypergeometric functions, that only depend on the momenta. We will separate its contributions into two classes according to the color ordering of Yang-Mills amplitudes. Each class contains three terms. The first class includes terms with color ordering $\{1,2,3,4,5,6\}, \{1,2,4,3,5,6\}, \{1,4,2,3,5,6\}$, whereas the second class includes terms with color ordering $\{1,3,2,4,5,6\},$ 
$\{1,3,4,2,5,6\}, \{1,4,3,2,5,6\}$. We will prove that in the soft limit $k_4 \rightarrow 0$, the sum of terms in the first class reduces to the soft factors multiplying $ A_{\rm YM}(1,2,3,5,6)F^{(2,3)}$, appearing in the five-point amplitude, and the sum of the terms in the second class reduces to the soft factors multiplying $ A_{\rm YM}(1,3,2,5,6) F^{(3,2)}$. It is convenient to solve for $\tilde{\lambda}_5$ and $\tilde{\lambda}_6$ using momentum conservation, and define
\bea
k'_5 &=& { |5 \rangle \langle 6| (1+2+3) \over \langle 56 \rangle }  \, ,  \quad
p_5 = { |5 \rangle \langle 6| 4 \over \langle 56 \rangle }  \, , \\
k'_6 &=& { |6 \rangle \langle 5| (1+2+3) \over \langle 65 \rangle }  \, ,  \quad
p_6 = { |6 \rangle \langle 5| 4 \over \langle 65 \rangle } \, .
\eea
Let us start with the terms in the first class. From the term with color ordering $\{1,2,3,4,5,6\}$, we have
\bea
F^{(234)}=-\int dz_2 dz_3 dz_4 \left( \prod_{i<l} |z_{il}|^{s_{il}} \right)
{s_{12} \over z_{12} } \left( {s_{34} \over z_{34} }  + {s_{35} \over z_{35} }  \right)
{s_{45} \over z_{45} }
\eea
here we use SL$(2)$ to fix $z_1=0, z_5=1$ and $z_6=\infty$. It is straightforward to see that this term produces a leading term given by 
\bea
{1 \over \delta^2 }S_{\rm YM}^{(0)}(345) F^{(2,3)}(1,2,3,5',6')  A_{\rm YM}(1,2,3,5',6') \, .
\eea
Focussing on the sub-leading part, we find 
\bea
\int dz_2 dz_3 \left( \prod_{i<l} |z_{il}|^{s_{il}} \right) {s_{12} \over z_{12}} F_{\delta}^{(234)}
\eea
where the Koba-Nielsen factor $\prod_{i<l} |z_{il}|^{s_{il}}$ is for five-point kinematics $\{ k_1, k_2, k_3, k'_5, k'_6 \}$ and $F_{\delta}^{(234)}$, of order $\mathcal{O}(\delta)$, is given by
\bea
F_{\delta}^{(234)} ={\delta \over z_{35} } \big[  
 ( s_{4'5'} + s_{34} + s_{3p_5} ) [1 + s_{35'} \log(1 - z_3)] + (s_{24}+ s_{2p_5} ) s_{35'} \log(1 - z_2) \big]  \, .  \nonumber
\eea
Similarly from the terms with color ordering $\{ 124356  \}$ and $\{ 142356  \}$, we find that the corresponding $F_{\delta}^{(243)}$ and $F_{\delta}^{(423)}$ are given by
\bea
F_{\delta}^{(243)} &=& \delta {s_{35'} \over z_{35} } \big[ s_{14} \log(z_{3}) + s_{24} \log(z_{23}) - s_{24} \log(1 - z_2) \big] \cr
F_{\delta}^{(423)} &=& - \delta {s_{35'} \over z_{35} }  s_{14} \log(z_3) \, .
\eea
Combining all the terms and putting back $\delta$-independent terms, we obtain 
\bea
{1 \over \delta^2 }A_{\rm YM}(1,2,3,5,6)\int dz_2 dz_3 {s_{12} \over z_{12}}\left( \prod_{i<l} |z_{il}|^{s_{il}} \right)
\left(
{ \langle 35\rangle \over \langle 34\rangle \langle 45\rangle  } F_{\delta}^{(234)}
+
{ \langle 23\rangle \over \langle 24\rangle \langle 43\rangle  } F_{\delta}^{(243)}
+{ \langle 12\rangle \over \langle 14\rangle \langle 42\rangle  } F_{\delta}^{(423)}
\right) \, , \nonumber
\eea
which we find to agree with 
\bea
{1 \over \delta } S_{\rm YM}^{(1)}(345) F^{(2,3)}(1,2,3,5',6') A_{\rm YM}(1,2,3,5',6') \, ,
\eea
at the level of the integrand. 

We then consider the expansion of terms in the second class. Firstly we observe that color orderings $\{ 132456  \}$ and $\{ 134256  \}$ both contain leading terms, and they combine to produce 
\bea
{1 \over \delta^2 } S_{\rm YM}^{(0)}(345)  F^{(3,2)}(1,3,2,5',6')A_{\rm YM}(1,3,2,5',6')  \, .
\eea 
Now consider the sub-leading terms. From color ordering $\{ 132456  \}$, we get 
\bea
\int dz_2 dz_3 \left( \prod_{i<l} |z_{il}|^{s_{il}} \right)  F_{\delta}^{(324)} { s_{13} \over z_{13} }
\eea
with the sub-leading term $ F_{\delta}^{(324)}$ is given by
\bea  \label{324}
F_{\delta}^{(324)}={s_{25'} \over z_{25} }\left[ ( s_{45'} + s_{34} + s_{3p_5} ) \log(1 - z_3) + (s_{24}+s_{2p_5} ) \log(1 - z_2) \right] + {1 \over z_{25}  } ( s_{24} + s_{2p_5}) \, .
\nonumber
\eea
Finally from terms with color ordering $\{ 134256  \}$ and $\{ 143256   \}$, we find
\bea
F_{\delta}^{(342)} &=& {\delta \over z_{25} } \left[s_{25'} \left( s_{24} \log(z_{32}) + k_2 \cdot p_5 \log(1- z_2) +   (s_{34} + s_{45'} + s_{3p_5} ) \log( 1 - z_3 ) \right)
+  s_{2p_5} \right] \cr
F_{\delta}^{(432)} &=& - {\delta \over z_{25} } s_{14} \left[ s_{25'}\log(z_3) + 1  \right]  \, .
\eea
Combining all the relevant terms, we find
\bea
{1 \over \delta^2 }A_{\rm YM}(1,3,2,5',6')\int dz_2 dz_3 {s_{13} \over z_{13}}\left( \prod_{i<l} |z_{il}|^{s_{il}} \right)
\left(
{ \langle 25\rangle \over \langle 24\rangle \langle 45\rangle  } F_{\delta}^{(324)}
+
{ \langle 32\rangle \over \langle 34\rangle \langle 42\rangle  } F_{\delta}^{(342)}
+{ \langle 13\rangle \over \langle 14\rangle \langle 43\rangle  } F_{\delta}^{(432)}
\right) \, , \nonumber
\eea
which can be checked to agree with 
\bea
{1 \over \delta }  S_{\rm YM}^{(1)}(345)  F^{(3,2)}(1,2,3,5',6') A_{\rm YM}(1,3,2,5',6') \, .
\eea
This ends the proof of the soft theorem for six-point open superstring amplitudes. 

\section{Soft theorem for five-point closed string amplitudes} \label{Appendix:5ptclose}
In this section we will check the validity of the soft theorem, especially $S^{(2)}_{\rm G}$, for closed superstring amplitudes at five points. As we discussed in section \ref{45ptclosed}, in order to use KLT formula, we need to expand five-point open superstring amplitudes to sub-sub-leading order. Here we will again solve for $\tilde{\lambda}_{4}$ and $\tilde{\lambda}_{5}$, and take $k_3$ to be the soft leg. Expanding up to order $\mathcal{O}(\delta^2)$, we obtain the five-point disk integral for open superstring amplitudes
\bea
F^{(2,3)} = { \Gamma(1 + s_{24'}) 
\Gamma(  1 + s_{12}) \over \Gamma(1 + s_{24'} + s_{12})} \left[ 1 + \delta f^{(2,3)}_1  + 
\delta^2 f^{(2,3)}_2  \right] + \mathcal{O}(\delta^3)
\eea
where the sub-leading and sub-sub-leading terms are given by 
\bea
f^{(2,3)}_1 &=& (s_{2p_4} + s_{23} + s_{34'}) \big[H(s_{24'}) - H(s_{24'} + s_{12}) \big] \, , \cr 
f^{(2,3)}_2 &=& 
 s_{13} s_{34'} \big[ ( \psi^{(0)}(s_{12}) -   \psi^{(0)}( 1 + s_{24'} + s_{12})) ( \psi^{(0)}(1 + s_{24'})  -   \psi^{(0)}(1 + s_{24'} + s_{12}))
 \cr
     &-&
   \psi^{(1)}( 1 + s_{24'} + s_{12}) 
    +
{     s_{12} \over (1 + s_{24'} + s_{12})}
       F(\{1, 1, 1, 1 + s_{12}\}, \{2, 2, 
       2 + s_{24'} + s_{12}\}, 1)  \big]  \cr
        &+&
    {1 \over 2}(s_{2p_4} + s_{23}+s_{34'})^2 \big[ 
     (\psi^{(0)}(1 + s_{24'} + s_{12})-\psi^{(0)}( 1 + s_{24'}))^2 + \psi^{(1)}(1 + s_{24'}) 
     \cr
        &-&
         \psi^{(1)}( 1 + s_{24'} + s_{12}) \big] -
           s_{34'} (s_{13}+s_{23}) \zeta_2 \, ,
\eea
where $H$ is the Harmonic Number, $F$ is the generalized hypergeometric function, and finally $\psi^{(m)}(z) = {d^{m+1} \over dz^{m+1} } \log(\Gamma(z)) $ is the PolyGamma function of order $m$. Similarly, we find the result of expanding $F^{(3,2)}$, which now starts from sub-leading order, 
\bea
F^{(3,2)} = s_{13} { \Gamma(1+ s_{24'}) \Gamma(1 + s_{12}) \over  \Gamma(1 + s_{24'} + s_{12}) } \big[ \delta f^{(3,2)}_1 + \delta^2 f^{(3,2)}_2 \big] + \mathcal{O}(\delta^3) \, ,
\eea
and $f^{(3,2)}_1, f^{(3,2)}_2$ that are given by
\bea
f^{(3,2)}_1 &=&  H(s_{24'} + s_{12}) - H(s_{12})   \, , \cr
f^{(3,2)}_2 &=& 
 s_{2p_4} \big[ (\psi^{(0)}(1 + s_{24'} + s_{12}) -\psi^{(0)}(s_{24'}) ) (\psi^{(0)}(1 + s_{12})  
    - 
 \psi^{(0)}(1 + s_{24'} + s_{12}))
\cr
  &+& \psi^{(1)}(    1 + s_{24'} + s_{12}) 
  -
  {1 \over s_{24'} } [\psi^{(0)}(1 + s_{12}) - \psi^{(0)}( 1 + s_{24'} + s_{12})] \big]
  \cr
  &+&    
 {  (1 + s_{12}) (s_{23} + s_{34'}) \over  1 + s_{24'} + s_{12} } 
       F(\{1, 1, 1, 2 + s_{12}\}, \{2, 2, 
      2 + s_{24'} + s_{12}\}, 1)
  \cr
      &-&
      {1 \over 2}  (s_{13} + s_{23}) 
      \big[ (\psi^{(0)}( 1 + s_{12})-  \psi^{(0)}( 1 + s_{24'} + s_{12}))^2 +  
      \psi^{(1)}( 1 + s_{12}) 
      \cr 
      &-&
 \psi^{(1)}(    1 + s_{24'} + s_{12}) \big] - (s_{23} + s_{34'}) \zeta_2   
       \, .
\eea
We thus obtain the expansion of the five-point open string amplitude up to sub-sub-leading order by substituting the above expansions into the expression for $\mathcal{A}_5(1,2,3,4,5)$,
\bea
\mathcal{A}_5(1,2,3,4,5) = 
F^{(2,3)} A_{\rm YM}(1,2,3,4,5) 
+ 
F^{(3,2)} A_{\rm YM}(1,3,2,4,5) \, . \nonumber
\eea
Similarly one can work out other open superstring amplitudes entering the KLT relation for the five-point closed superstring amplitude,
\ba
\mathcal{M}_5(\{1,2,3,4,5\})&=& \pi^{-2} \big( \mathcal{A}_5(1,2,3,4,5) \mathcal{A}_5(1,4,3,5,2)\sin(\pi s_{12}) \sin(\pi s_{34})\nl
 &+& \mathcal{A}_5(5,1,3,2,4) \mathcal{A}_5(2,5,3,1,4)\sin(\pi s_{13}) \sin(\pi s_{24}) \big)\, .
\ea
With the above results up to the necessary order, we find that $\mathcal{M}_5(\{1,2,3,4,5\})$ satisfies the soft theorem by numerically comparing it with 
\bea
\left( {1 \over \delta^3} S_{\rm G}^{(0)}(3)  + {1 \over \delta^2} S_{\rm G}^{(1)}(3) + {1 \over \delta} S_{\rm G}^{(2)}(3) \right) \mathcal{M}_4(\{1,2,4',5'\}) \, . 
\eea
This explicit numerical test of the soft theorem is consistent with the argument based on BCFW recursion relations and the world-sheet OPE analysis presented in Sections \ref{string1} and \ref{stringOPE}.

\end{document}